\pdfoutput=1
\documentclass[twocolumn, showpacs,amsmath,amssymb,floatfix,aps, prl, 10pt]{revtex4-1}
\usepackage[latin1]{inputenc}
\usepackage[american]{babel}
\usepackage[T1]{fontenc}
\usepackage{mathrsfs}
\usepackage{hyperref}
\usepackage{multirow}
\usepackage{bm, epsfig}
\usepackage{graphicx}
\usepackage{subeqnarray}
\usepackage{upgreek}
\usepackage[normalem]{ulem}
\setcitestyle{super}
\usepackage{helvet}
\usepackage{afterpage}

\hypersetup{colorlinks=true, urlcolor=blue, citecolor=blue, filecolor=blue, linkcolor=blue}
\newcommand*{\diff}{\mathop{}\!\mathrm{d}}

\newcommand*{\Imm}{\mathop{}\!\mathbf{Im}}
\newcommand{\uimm}{\mathrm{i}}
\newcommand{\eu}{\mathrm{e}}

\newcommand{\unit}[1]{\ensuremath{\, \mathrm{#1}}}

\DeclareSymbolFont{bbold}{U}{bbold}{m}{n}
\DeclareSymbolFontAlphabet{\mathbbold}{bbold}

\begin{document}

\title{Broadband high-resolution x-ray frequency combs}

\date{\today}

% \author{Stefano M. Cavaletto$^{*}$, Zolt\'an Harman, Christian Ott, Christian Buth$^{\dagger}$,Thomas Pfeifer \& Christoph H. Keitel}
% 
% \begin{affiliations}
%  \item Max-Planck-Institut f\"{u}r Kernphysik, Saupfercheckweg~1, 69117~Heidelberg, Germany. *e-mail: smcavaletto@gmail.com $\dagger$ Present address: Max-Planck-Institut f\"ur Quantenoptik, Hans-Kopfermann-Stra\ss{}e~1, 85748~Garching bei M\"unchen, Germany
% \end{affiliations}

\author{Stefano M. Cavaletto}
\email[Corresponding author. Email: ]{smcavaletto@gmail.com}
\affiliation{Max-Planck-Institut f\"{u}r Kernphysik, Saupfercheckweg~1, 69117~Heidelberg, Germany}
\author{Zolt\'an Harman}
\affiliation{Max-Planck-Institut f\"{u}r Kernphysik, Saupfercheckweg~1, 69117~Heidelberg, Germany}
\author{Christian Ott}
\affiliation{Max-Planck-Institut f\"{u}r Kernphysik, Saupfercheckweg~1, 69117~Heidelberg, Germany}
\author{Christian Buth}
\altaffiliation[Present address: ]{Max-Planck-Institut f\"ur Quantenoptik, Hans-Kopfermann-Stra\ss{}e~1, 85748~Garching bei M\"unchen, Germany}
\affiliation{Max-Planck-Institut f\"{u}r Kernphysik, Saupfercheckweg~1, 69117~Heidelberg, Germany}
\author{Thomas Pfeifer}
\affiliation{Max-Planck-Institut f\"{u}r Kernphysik, Saupfercheckweg~1, 69117~Heidelberg, Germany}
\author{Christoph H. Keitel}
\affiliation{Max-Planck-Institut f\"{u}r Kernphysik, Saupfercheckweg~1, 69117~Heidelberg, Germany}

% Meaning of the PACS numbers: phraseas
% 32.80.Qk: Coherent control of atomic interactions with photons
% 32.50.+d: Fluorescence, phosphorescence (including quenching)
% 42.50.Ct: Quantum description of interaction of light and matter; related experiments
% 42.62.Eh: Metrological applications; optical frequency synthesizers for precision spectroscopy (see also 06.20.-f Metrology in metrology, measurements, and laboratory procedures)

\maketitle

% -------------------------------------------------------------------------------------------------------------------------
% -------------------------------------------------------------------------------------------------------------------------

%\sout{This is a deleted or incorrect sentence.} 

{\small \textbf{\textsf{Optical frequency combs have had a remarkable impact on precision spectroscopy\cite{Nature.416.233, RevModPhys.75.325,Diddams:10}. Enabling this technology in the x-ray domain is expected to result in wide-ranging applications, such as stringent tests of astrophysical models and quantum electrodynamics\cite{BernittNature}, a more sensitive search for the variability of fundamental constants\cite{PhysRevLett.106.210802}, and precision studies of nuclear structure\cite{PhysRevLett.96.142501}. Ultraprecise x-ray atomic clocks may also be envisaged\cite{PhysRevLett.104.200802}. In this work, an \mbox{x-ray} pulse-shaping method is put forward to generate a comb in the absorption spectrum of an ultrashort high-frequency pulse. The method employs an {optical-frequency-comb} laser, manipulating the system's dipole response to imprint a comb on { an excited transition with a high photon energy}. The described scheme provides higher comb frequencies and requires lower optical-comb peak intensities than currently explored methods\cite{PhysRevLett.94.193201, Nature.436.234, Nature.482.68}, preserves the overall width of the optical comb, and may be implemented by {presently} available x-ray technology\cite{0022-3727-45-21-213001}.}}}

%otherwise not directly accessible

The spectrum of an optical frequency comb consists of equally spaced, precisely known peaks, centred at an optical frequency\cite{Nature.416.233, RevModPhys.75.325}. Optical-comb generation was initially pursued via intracavity phase modulation\cite{250392,Ye:97} and subsequently with stabilized mode-locked lasers \cite{Jones28042000}. Optical frequency combs are employed\cite{Diddams:10}, e.g., in precision spectroscopy\cite{PhysRevLett.82.3568}, all-optical atomic clocks\cite{Diddams03082001}, astronomical spectrographs calibration \cite{Steinmetz05092008}, attosecond science\cite{Nature.421.611}, and control of atomic coherence\cite{PhysRevLett.96.153001}% and molecular dynamics\cite{PhysRevLett.98.113004}
. \mbox{X-ray} frequency combs would {enable} the aforementioned applications in the \mbox{x-ray} range. Stringent tests of fundamental physics may be pursued, e.g., accurate measurements of transition energies in highly charged ions, which are predicted to be more sensitive to the variability of fundamental constants\cite{PhysRevLett.106.210802} than currently investigated species. 

Presently, XUV combs ($\sim30\unit{eV}$) are generated\cite{Nature.482.68} via intracavity high-order harmonic generation (HHG)\cite{PhysRevLett.94.193201, Nature.436.234}. %
% This would lead to a plethora of applications, such as structure probe in atomic nuclei\cite{}, more stringent tests of quantum electrodynamics and astrophysical models\cite{}, and search for variability of fundamental constants in highly charged ions\cite{}. Ultraprecise x-ray atomic clocks may also be envisioned\cite{}. The generation of XUV combs via intracavity HHG\cite{PhysRevLett.94.193201, Nature.436.234} is presently recognized as the most promising tool to extend precision spectroscopy to the short-wavelength domain \cite{Nature.482.68}. 
A femtosecond enhancement cavity is utilized to reach the required peak intensities\cite{PhysRevLett.94.193201, Nature.436.234, Nature.482.68}, up to $\sim$\,$10^{14}\,\unit{W/cm^2}$. %Efficiency limitations of HHG owing to relativistic effects at high harmonic orders \cite{Kohler2012159} render such solution not advisable for \mbox{x-ray} frequencies. Investigation of schemes to overcome these challenges are therefore timely.
However, relativistic effects limit the efficiency of HHG at high harmonic orders \cite{Kohler2012159}. The investigation of schemes to further {increase the carrier frequency of the comb} at accessible driving intensities is therefore required.

\begin{figure}[tb]
\centering%
\includegraphics[width=\linewidth, keepaspectratio]{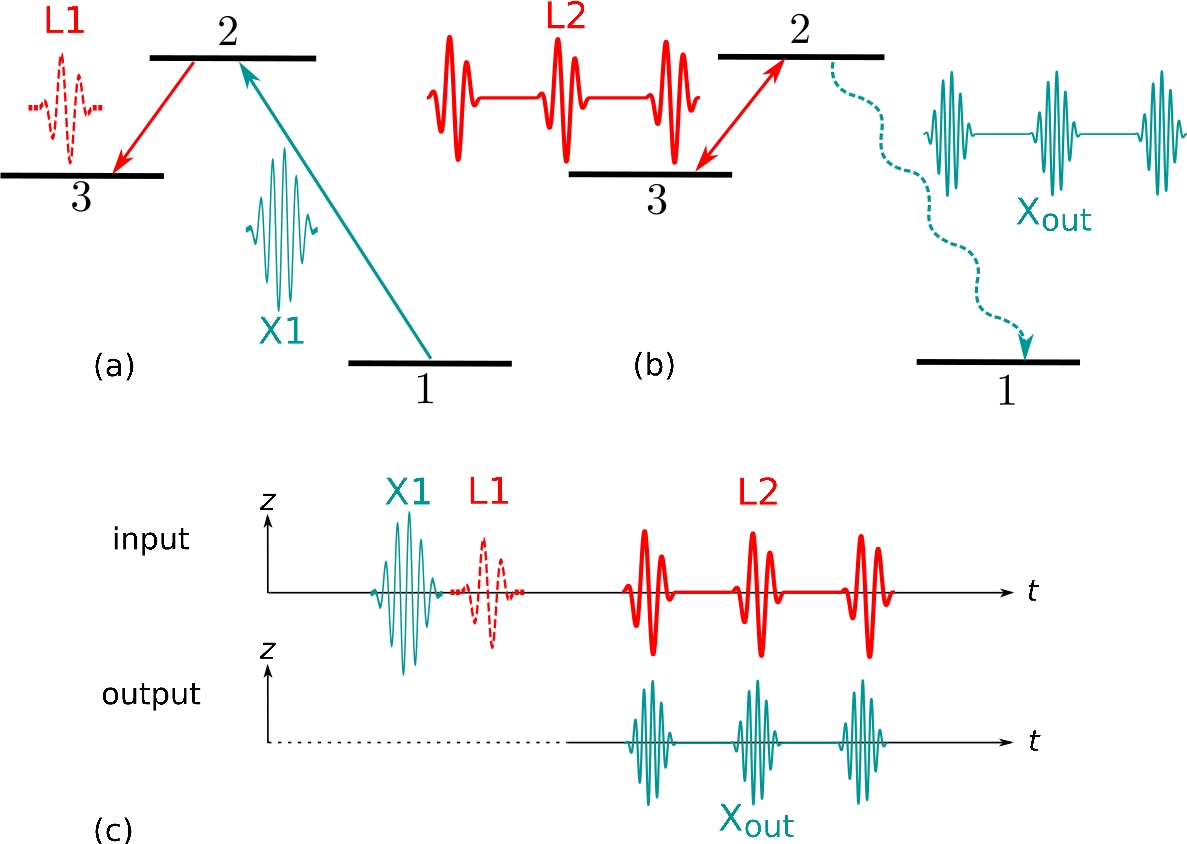}
\caption{\textbf{Three-level scheme used to describe the interaction between the model system and the driving fields}. 
(a) A low-density ensemble of ions, modelled as a three-level system, is driven by an ultrashort, broadband \mbox{x-ray} pulse (X1, solid, blue), exciting the fast decaying level 2, followed by an optical pulse (L1, dashed, red) coupling this excited state to the metastable state 3. Thereby the system is prepared in an initial state which is a superposition of states 1 and 3. (b) An optical frequency comb (L2, solid, red) is subsequently used to periodically drive the optical transition $2\leftrightarrow 3$. {The emitted x~rays ($\mathrm{X_{out}}$, dashed, wavy, blue) lead to either gain or attenuation of the incident pulse X1 while it propagates through the medium. This is detected via the absorption spectrum of the transmitted pulse X1.} (c) Input and output pulses from the previous panels. The pulses are linearly polarized {along the $z$~direction} and copropagating.}%
\label{fig:Scheme}
\end{figure}

%Alternative ideas may come from the rapidly growing field of \mbox{x-ray} quantum optics\cite{doi:10.1080/09500340.2012.752113}. 
Short-wavelength light sources with improved brilliance and bandwidth\cite{0022-3727-45-21-213001} enable {studies} of \mbox{x-ray} quantum optics, e.g., in highly charged ions or nuclei \cite{BernittNature,PhysRevLett.96.142501, RoehlsbergNature}. Recently, an amplitude-shaping scheme was put forward to imprint a comb onto narrowband x~rays\cite{PhysRevA.88.063402}. Comb generation was also suggested\cite{Zuoye} via quantum phase modulation\cite{Ott10052013}. However, these schemes are conditioned either by demanding requirements on the \mbox{x-ray} source bandwidth\cite{PhysRevA.88.063402}, or by the spectral width of the emerging comb\cite{Zuoye}. In contrast, {here, the} predicted comb requires \mbox{x-ray} pulses as presently provided by free-electron lasers\cite{0022-3727-45-21-213001} (FELs) and, being as wide as the employed optical frequency comb, is suitable to bridge the gap between a reference level and a nearby unknown level\cite{250392,Ye:97}.

{The key idea of our method is introduced via the three-level system of Fig.~\ref{fig:Scheme}; we then prove its viability in a realistic atomic implementation. The excited level of the high-energy transition $1\rightarrow 2$} is coupled by an optical laser to a third metastable level. This is chosen such that it cannot be excited in a one-photon transition and decays via two-photon emission. Hence, the decay rate $\varGamma_3$ of this dark state is by orders of magnitude lower than the decay rate $\varGamma_2$ of the bright (fast decaying) level 2. %The third transition $1\rightarrow 3$ is not electric-dipole allowed and both driving and spontaneous decay are suppressed by orders of magnitude. State $i$ has energy $\omega_i$ and the energy between the states $i$ and $j$ is given by $\omega_{ij} = \omega_i - \omega_j$, $i,\,j\in\{1,\,2,\,3\}$. The excited state 2 decays back to the ground state with the decay rate $\varGamma_2$, which in our model is by orders of magnitude larger than the decay rate $\varGamma_3$ of state 3. 
The system is described via the density operator $\hat{\rho}(t)$, of matrix elements $\rho_{ij}(t)$, with $i,\,j\in\{1,\,2,\,3\}$. The diagonal element $\rho_{ii}(t)$ describes the occupation probability of level $i$, whereas the off-diagonal element $\rho_{ij}(t)$ describes the coherence between states $i$ and $j$. Atomic units are used throughout unless otherwise stated.

As shown in Fig.~\ref{fig:Scheme}, an \mbox{x-ray} pulse (X1) excites the atomic system, whose dipole response is given by the coherence $\rho_{12}(t)$. While propagating through the medium, the pulse interferes with the radiation $\mathrm{X_{out}}$ emitted by the driven system, thus creating gain or attenuation in the absorption spectrum
\begin{equation}
\sigma(\omega) \propto -\omega \Imm\{\tilde{\rho}_{12}(\omega)\}
\label{eq:transabsspec}
\end{equation}
of the transmitted pulse X1, with the frequency-dependent dipole response\cite{doi:10.1146/annurev.pc.43.100192.002433} $\tilde{\rho}_{12}(\omega) = \int_{-\infty}^{\infty}\rho_{12}(t)\,\eu^{-\uimm\omega t}\,\diff t$. Here, $\sigma(\omega)>0$ corresponds to absorption and $\sigma(\omega)<0$ to gain. 

If only the pulse X1 is employed, {spontaneous decay of the bright level follows its \mbox{x-ray} excitation}. The resulting absorption spectrum, centred at the transition energy $\omega_{21}$ between states 2 and 1, exhibits a Lorentzian profile of width $\varGamma_2$. {Based on previous experiments on time-domain control\cite{Marian17122004, Ott10052013}, we} augment this x-ray-only scheme by two optical fields [Figs.~\ref{fig:Scheme}(a) and (b)]. The first optical pulse [L1 in Fig.~\ref{fig:Scheme}(a)] depletes the fast decaying state 2 by transferring coherences and population into the {metastable level 3. By preparing the system in a long-lived initial state which is a superposition of states 1 and 3, the fast decay of the system due to the large decay rate of level 2 is circumvented (see the Supplemental Information)}. An optical frequency comb [L2 in Fig.~\ref{fig:Scheme}(b)] is subsequently employed to periodically modulate the emission out of the bright state, leading to a pulse train at the \mbox{x-ray} transition $\omega_{21}$.

To better understand this, we analyse the dynamics of the system interacting with the optical pulse train L2. The electric field $\boldsymbol{\mathcal{E}}_{\mathrm{L2}}(t) = \bar{\mathcal{E}}_{\mathrm{L2}}(t) \cos(\omega_{\mathrm{L2}}t)\hat{\boldsymbol{e}}_z$ has envelope $\bar{\mathcal{E}}_{\mathrm{L2}}(t)$ and carrier frequency $\omega_{\mathrm{L2}}$. The linear polarization $\hat{\boldsymbol{e}}_z$, parallel to the dipole-moment matrix element $\boldsymbol{d}_{23}$ between states 2 and 3, leads to the instantaneous Rabi frequency\cite{Scully:QuantumOptics} $\varOmega_{\mathrm{R}}(t) = \boldsymbol{d}_{23}\cdot \hat{\boldsymbol{e}}_z\,\bar{\mathcal{E}}_{\mathrm{L2}}(t)$. {The envelope $\bar{\mathcal{E}}_{\mathrm{L2}}(t)$ consists of identical pulses, located at $t_j = t_0 + j T_{\mathrm{p}}$, $j\in \mathbb{N}_0$, with repetition period $T_{\mathrm{p}}$ and single-pulse FWHM duration $\tau\ll T_{\mathrm{p}}$.} The peak intensity $I$ corresponds to a maximum field strength $\bar{\mathcal{E}}_{\mathrm{max}} = \sqrt{8\pi\alpha I}$, with $\alpha$ being the fine-structure constant. By integrating the corresponding Rabi frequency $\varOmega_{\mathrm{R}}(t)$ over $T_{\mathrm{p}}$, we define the pulse area $Q \propto \bar{\mathcal{E}}_{\mathrm{max}}\,\tau$ (Supplemental Information). The coupled differential equations\cite{Scully:QuantumOptics}
\begin{subeqnarray}
\tfrac{\diff \rho_{12}}{\diff t} &=  \bigl(\uimm\,\omega_{21} - \tfrac{\varGamma_{2}}{2} \bigr)\rho_{12}(t) + \uimm \tfrac{\varOmega_{\mathrm{R}}(t)}{2}\,\eu^{\uimm \omega_{\mathrm{L}}t}\,\rho_{13}(t),\\
\tfrac{\diff \varrho_{13}}{\diff t} &= \bigl(\uimm\, \omega_{31} - \tfrac{\varGamma_{3}}{2}\bigr) \rho_{13}(t) + \uimm \tfrac{\varOmega_{\mathrm{R}}(t)}{2}\,\eu^{-\uimm\omega_{\mathrm{L}}t}\,\rho_{12}(t),
\label{eq:firstorderequation}
\end{subeqnarray} 
describe the dynamical evolution of the system, where the coherence $\rho_{13}(t)$ between states 1 and 3 decays with rate $\varGamma_3\ll\varGamma_2$. 

\begin{figure}[t]
\centering%
\includegraphics[width=0.96\linewidth, keepaspectratio]{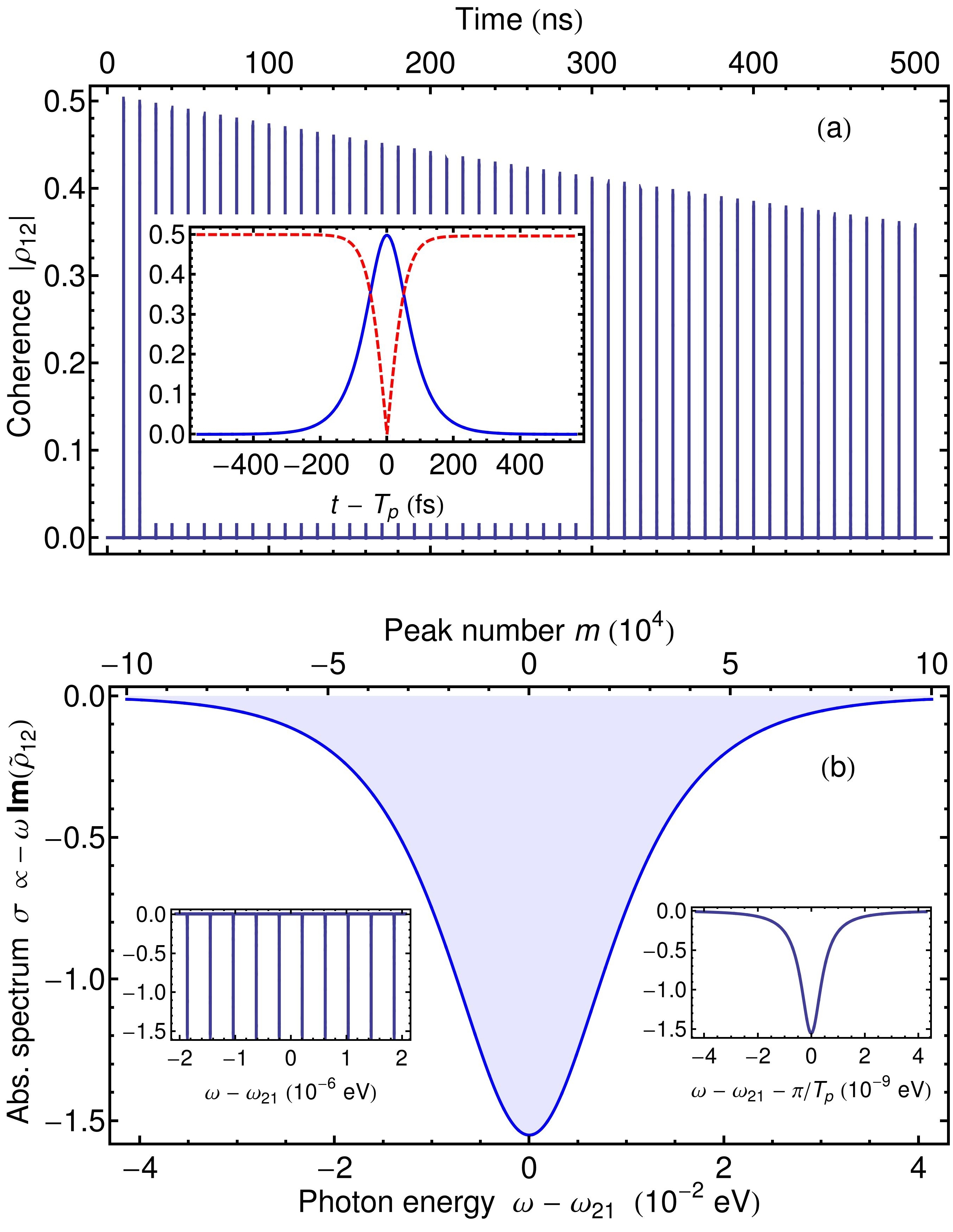}
\caption{\textbf{Time evolution and \mbox{x-ray} frequency comb absorption spectrum driven by a resonant optical frequency comb.} The three-level system of Fig.~\ref{fig:Scheme} is used to describe $\mathrm{Be}^{2+}$ ions driven by a train of optical pulses tuned to the $2\leftrightarrow 3$ transition. The driving comb is modelled via present-day parameters\cite{Nature.482.68}, i.e., a single-pulse FWHM duration of $100\,\unit{fs}$, a spectral width of $0.07\,\,\mathrm{eV}$, and 100-MHz repetition frequency. The peak intensity of $2.57\times 10^{10}\unit{W/cm^2}$ corresponds to a $2\pi$ pulse area. Panel (a) displays the time evolution of $|\rho_{12}(t)|$ during the first 50 pulses of the optical frequency comb. The inset shows $|\rho_{12}(t)|$ (blue, solid line) and $|\rho_{13}(t)|$ (red, dashed line) in the presence of the first pulse. Panel (b) exhibits the peak amplitudes of the frequency comb featured by the absorption spectrum, normalized to the maximum of the single-peak Lorentzian spectrum in the absence of optical control. The comb is centred at the \mbox{x-ray} transition energy $\omega_{21} = 123.7\unit{eV}$. The comb structure is apparent from the inset on the left, while the inset on the right shows the shape of a single comb tooth. {For an \mbox{x-ray} pulse\cite{0022-3727-45-21-213001} tuned to the transition energy $\omega_{21}$, with duration of $100\,\mathrm{fs}$, beam radius of $50\,\mathrm{\mu m}$, peak intensity of $2.5\times10^{12}\,\mathrm{W/cm^2}$, and bandwidth of $1\,\mathrm{eV}$, driving a sample of ions with density\cite{BernittNature} of $10^9\,\mathrm{cm^{-3}}$ over a length of $2 \,\mathrm{cm}$, this corresponds to a power per comb line of $\sim\,30\,\mathrm{pW}$ (Supplemental Information), on the same order of magnitude as \mbox{XUV} combs generated via HHG\cite{Nature.482.68}.}}%
\label{fig:Showtuned}
\end{figure}

By interacting with the fields X1 and L1 [Fig.~\ref{fig:Scheme}(a)], the system is prepared in an initial state in which the coherence is ``stored'' in the slowly decaying function $\rho_{13}(t)$, while $\rho_{12}(t)$ initially vanishes. Then, the train of pulses L2, with properly set peak intensity and duration, is employed to periodically transfer the coherence to the fast decaying function $\rho_{12}(t)$. {Specifically}, by either using $2\pi$-area pulses, or pulses detuned from the corresponding optical transition, {we show in the Supplemental Information that $\rho_{12}$ can be converted into a vanishing value at the end of each optical pulse\cite{PhysRevA.88.063402}.} Thereby, the envelope of $\rho_{12}$ is shaped into a function closely following $\bar{\mathcal{E}}_{\mathrm{L2}}(t)$. A train of short pulses ($\mathrm{X_{out}}$) is thus emitted at the \mbox{x-ray} transition energy $\omega_{21}$ and, from Eq.~(\ref{eq:transabsspec}), a wide frequency comb appears in the absorption spectrum of the transmitted pulse X1. 

\begin{figure}[t]
\centering%
\includegraphics[width=0.96\linewidth, keepaspectratio]{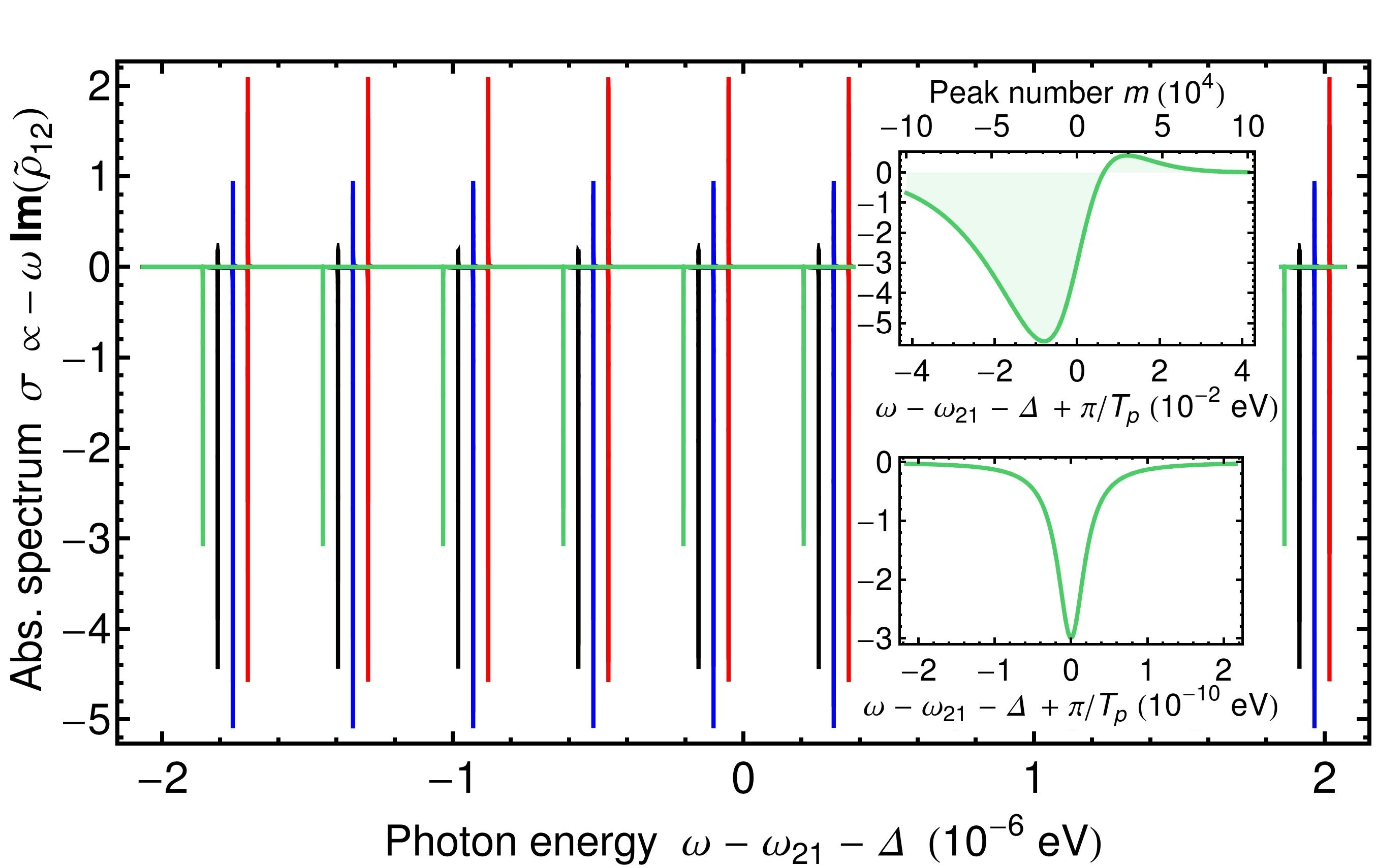}
\caption{\textbf{\mbox{X-ray} comb absorption spectrum driven by a detuned optical frequency comb.} We use \mbox{Be$^{2+}$} ions to implement the three-level scheme of Fig.~\ref{fig:Scheme}. A train of off-resonant, 1.5-eV pulses is employed, with a detuning of $\varDelta = -0.52\,\unit{eV}$. Single-pulse FWHM duration and repetition period are set as in Fig.~\ref{fig:Showtuned}. The spectra associated with different values of the peak intensity, i.e., $4.54\times 10^{11}\unit{W/cm^2}$ (red), $9.14\times 10^{11}\unit{W/cm^2}$ (blue), $1.38\times 10^{12}\unit{W/cm^2}$ (black), and $1.85\times 10^{12}\unit{W/cm^2}$ (green), are displayed{, normalized to the peak of the Lorentzian spectrum in the absence of optical control}. For the green curve, the insets show the amplitude of the peaks in the comb (top) and the shape of a single comb tooth (bottom).}%
\label{fig:Showdetuned}
\end{figure}

To test the feasibility of our scheme, we apply the three-level system of Fig.~\ref{fig:Scheme} to model isolated transitions in heliumlike $\mathrm{Be}^{2+}$. The transition energy\cite{PhysRevA.81.022507} of $123.7\,\mathrm{eV}$ between $1s^2\,^1S_0$ and $1s\,2p\,^1P_1$ (levels 1 and 2) is in the \mbox{x-ray} range, while an optical energy of $2.02\,\mathrm{eV}$ separates $1s\,2s\,^1S_0$ (level 3) from the bright state. The two-photon decay rate\cite{PhysRevA.56.1288} $\varGamma_{3} = 1.8\times 10^4\unit{s^{-1}}$ is negligible compared to $\varGamma_{2} = 1.2\times 10^{11}\unit{s^{-1}}$, as computed with \texttt{grasp2K}\cite{DBLP:journals/cphysics/JonssonHFG07}. The three levels are non-autoionizing and photoionization can be neglected at the intensities discussed here. 
%Present-day parameters are used to model the optical frequency comb: we assume a single-pulse FWHM $T_{\mathrm{FWHM}} = 100\,\unit{fs}$, corresponding to a bandwidth $\gamma = 0.01\unit{eV}$. The repetition period is $T_{\mathrm{p}} = 10\,\unit{ns}$, $1/T_{\mathrm{p}} = 100\,\unit{MHz}$ \cite{Nature.482.68}, i.e., $2\pi/T_{\mathrm{p}} = 4.1\times 10^{-7}\,\unit{eV}$.

In Fig.~\ref{fig:Showtuned}, we employ an optical frequency comb tuned to the optical transition. The peak intensity of $2.57\times10^{10}\unit{W/cm^2}$, corresponding to $2\pi$ pulses, is by {several} orders of magnitude lower than the intensity required for HHG-based schemes\cite{PhysRevLett.94.193201, Nature.436.234, Nature.482.68}. As shown in Fig.~\ref{fig:Showtuned}(a), the coherence, stored in $\rho_{13}(t)$ in the long time interval in between two pulses, is transferred to $\rho_{12}(t)$ only in the presence of an optical pulse, with corresponding decay of the bright state. This {results in} the \mbox{x-ray} comb of Fig.~\ref{fig:Showtuned}(b), centred and locked to the natural transition frequency at $123.7\unit{eV}$, spanning as wide as the driving optical frequency comb, and {consisting} of Lorentzian peaks separated by the optical-comb repetition frequency of $100\unit{MHz}$ ($4.1\times 10^{-7}\,\unit{eV}$). The width of {these} peaks, given by the inverse of the long, effective decay time of the system, is by orders of magnitude smaller than the natural width of the bright state. This leads to a correspondingly high spectral resolution and attainable precision of the comb.

In Fig.~\ref{fig:Showdetuned}, the scheme is used to analyse off-resonant optical driving. We employ pulses with $1.5$-$\mathrm{eV}$ photon energy from mode-locked Ti:sapphire lasers. The chosen peak intensities, near $10^{12}\,\mathrm{W/cm^2}$, are lower than those presently used for XUV-comb generation with HHG\cite{PhysRevLett.94.193201, Nature.436.234, Nature.482.68}. Position and line shape of the peaks in the absorption spectrum depend controllably on the peak intensity (see the Supplemental Information). Despite its asymmetric shape, the \mbox{x-ray} absorption spectrum is as wide as the optical frequency comb.

{The scheme is, in principle, applicable to atomic systems at higher \mbox{x-ray} energies, provided an \mbox{x-ray} source\cite{0022-3727-45-21-213001} is available at that photon energy and an optical transition can be identified between the bright state and a metastable level. Furthermore, the duration of the optical pulses and the time delay between X1 and L1 have to be smaller than the decay time of the excited state, to enable the preparation of the system and its further manipulation [see Fig.~\ref{fig:Scheme} and the Supplemental Information].}

In conclusion, we developed a scheme to %control the emission of x~rays from a bright level by coupling it to a nearby dark state with an optical frequency comb. Thereby, 
{imprint} a frequency comb onto the absorption spectrum of an \mbox{x-ray} {FEL} pulse%such as those currently available at FELs
\cite{0022-3727-45-21-213001}. The spectrum consists of equally spaced lines whose exact positions, referenced to an atomic transition, may be used to precisely bridge a reference level and an unknown \mbox{x-ray} frequency%{, thus facilitating applications of frequency-comb technology at high photon energies}
{.} Furthermore, by using transitions with a higher energy, e.g., $1s^2\rightarrow1s \,np$ in heliumlike ions of higher nuclear charges, or nuclear transitions, {the presented} scheme may enable comb generation up to gamma frequencies.

% \acknowledgments
% The authors declare no competing financial interests. Supplementary information accompanies this paper. Correspondence and requests for materials should be addressed to S.M.C.

\bibliographystyle{naturemag}

\begin{thebibliography}{10}
\expandafter\ifx\csname url\endcsname\relax
  \def\url#1{\texttt{#1}}\fi
\expandafter\ifx\csname urlprefix\endcsname\relax\def\urlprefix{URL }\fi
\providecommand{\bibinfo}[2]{#2}
\providecommand{\eprint}[2][]{\url{#2}}

\bibitem{Nature.416.233}
\bibinfo{author}{Udem, T.}, \bibinfo{author}{Holzwarth, R.} \&
  \bibinfo{author}{H\"ansch, T.~W.}
\newblock \bibinfo{title}{Optical frequency metrology}.
\newblock \emph{\bibinfo{journal}{Nature}} \textbf{\bibinfo{volume}{416}},
  \bibinfo{pages}{233--237} (\bibinfo{year}{2002}).
\newblock \urlprefix\url{http://dx.doi.org/10.1038/416233a}.

\bibitem{RevModPhys.75.325}
\bibinfo{author}{Cundiff, S.~T.} \& \bibinfo{author}{Ye, J.}
\newblock \bibinfo{title}{\textit{Colloquium} : Femtosecond optical frequency
  combs}.
\newblock \emph{\bibinfo{journal}{Rev. Mod. Phys.}}
  \textbf{\bibinfo{volume}{75}}, \bibinfo{pages}{325--342}
  (\bibinfo{year}{2003}).
\newblock \urlprefix\url{http://link.aps.org/doi/10.1103/RevModPhys.75.325}.

\bibitem{Diddams:10}
\bibinfo{author}{Diddams, S.~A.}
\newblock \bibinfo{title}{{The evolving optical frequency comb [Invited]}}.
\newblock \emph{\bibinfo{journal}{J. Opt. Soc. Am. B}}
  \textbf{\bibinfo{volume}{27}}, \bibinfo{pages}{B51--B62}
  (\bibinfo{year}{2010}).
\newblock
  \urlprefix\url{http://josab.osa.org/abstract.cfm?URI=josab-27-11-B51}.

\bibitem{BernittNature}
\bibinfo{author}{Bernitt, S.} \emph{et~al.}
\newblock \bibinfo{title}{{An unexpectedly low oscillator strength as the
  origin of the Fe{\thinspace}XVII emission problem}}.
\newblock \emph{\bibinfo{journal}{Nature}} \textbf{\bibinfo{volume}{492}},
  \bibinfo{pages}{225--228} (\bibinfo{year}{2012}).
\newblock \urlprefix\url{http://dx.doi.org/10.1038/nature11627}.

\bibitem{PhysRevLett.106.210802}
\bibinfo{author}{Berengut, J.~C.}, \bibinfo{author}{Dzuba, V.~A.},
  \bibinfo{author}{Flambaum, V.~V.} \& \bibinfo{author}{Ong, A.}
\newblock \bibinfo{title}{Electron-hole transitions in multiply charged ions
  for precision laser spectroscopy and searching for variations in $\alpha$}.
\newblock \emph{\bibinfo{journal}{Phys. Rev. Lett.}}
  \textbf{\bibinfo{volume}{106}}, \bibinfo{pages}{210802}
  (\bibinfo{year}{2011}).
\newblock
  \urlprefix\url{http://link.aps.org/doi/10.1103/PhysRevLett.106.210802}.

\bibitem{PhysRevLett.96.142501}
\bibinfo{author}{B\"urvenich, T.~J.}, \bibinfo{author}{Evers, J.} \&
  \bibinfo{author}{Keitel, C.~H.}
\newblock \bibinfo{title}{Nuclear quantum optics with x-ray laser pulses}.
\newblock \emph{\bibinfo{journal}{Phys. Rev. Lett.}}
  \textbf{\bibinfo{volume}{96}}, \bibinfo{pages}{142501}
  (\bibinfo{year}{2006}).
\newblock
  \urlprefix\url{http://link.aps.org/doi/10.1103/PhysRevLett.96.142501}.

\bibitem{PhysRevLett.104.200802}
\bibinfo{author}{Rellergert, W.~G.} \emph{et~al.}
\newblock \bibinfo{title}{Constraining the evolution of the fundamental
  constants with a solid-state optical frequency reference based on the
  $^{229}\mathrm{Th}$ nucleus}.
\newblock \emph{\bibinfo{journal}{Phys. Rev. Lett.}}
  \textbf{\bibinfo{volume}{104}}, \bibinfo{pages}{200802}
  (\bibinfo{year}{2010}).
\newblock
  \urlprefix\url{http://link.aps.org/doi/10.1103/PhysRevLett.104.200802}.

\bibitem{PhysRevLett.94.193201}
\bibinfo{author}{Jones, R.~J.}, \bibinfo{author}{Moll, K.~D.},
  \bibinfo{author}{Thorpe, M.~J.} \& \bibinfo{author}{Ye, J.}
\newblock \bibinfo{title}{Phase-coherent frequency combs in the vacuum
  ultraviolet via high-harmonic generation inside a femtosecond enhancement
  cavity}.
\newblock \emph{\bibinfo{journal}{Phys. Rev. Lett.}}
  \textbf{\bibinfo{volume}{94}}, \bibinfo{pages}{193201}
  (\bibinfo{year}{2005}).
\newblock
  \urlprefix\url{http://link.aps.org/doi/10.1103/PhysRevLett.94.193201}.

\bibitem{Nature.436.234}
\bibinfo{author}{Gohle, C.} \emph{et~al.}
\newblock \bibinfo{title}{A frequency comb in the extreme ultraviolet}.
\newblock \emph{\bibinfo{journal}{Nature}} \textbf{\bibinfo{volume}{436}},
  \bibinfo{pages}{234--237} (\bibinfo{year}{2005}).
\newblock \urlprefix\url{http://dx.doi.org/10.1038/nature03851}.

\bibitem{Nature.482.68}
\bibinfo{author}{Cing\"oz, A.} \emph{et~al.}
\newblock \bibinfo{title}{Direct frequency comb spectroscopy in the extreme
  ultraviolet}.
\newblock \emph{\bibinfo{journal}{Nature}} \textbf{\bibinfo{volume}{482}},
  \bibinfo{pages}{68--71} (\bibinfo{year}{2012}).
\newblock \urlprefix\url{http://dx.doi.org/10.1038/nature10711}.

\bibitem{0022-3727-45-21-213001}
\bibinfo{author}{Rebernik~Ribic, P.} \& \bibinfo{author}{Margaritondo, G.}
\newblock \bibinfo{title}{{Status and prospects of x-ray free-electron lasers
  (X-FELs): a simple presentation}}.
\newblock \emph{\bibinfo{journal}{Journal of Physics D: Applied Physics}}
  \textbf{\bibinfo{volume}{45}}, \bibinfo{pages}{213001}
  (\bibinfo{year}{2012}).
\newblock \urlprefix\url{http://stacks.iop.org/0022-3727/45/i=21/a=213001}.

\bibitem{250392}
\bibinfo{author}{Kourogi, M.}, \bibinfo{author}{Nakagawa, K.} \&
  \bibinfo{author}{Ohtsu, M.}
\newblock \bibinfo{title}{Wide-span optical frequency comb generator for
  accurate optical frequency difference measurement}.
\newblock \emph{\bibinfo{journal}{Quantum Electronics, IEEE Journal of}}
  \textbf{\bibinfo{volume}{29}}, \bibinfo{pages}{2693--2701}
  (\bibinfo{year}{1993}).

\bibitem{Ye:97}
\bibinfo{author}{Ye, J.}, \bibinfo{author}{Ma, L.-S.}, \bibinfo{author}{Daly,
  T.} \& \bibinfo{author}{Hall, J.~L.}
\newblock \bibinfo{title}{Highly selective terahertz optical frequency comb
  generator}.
\newblock \emph{\bibinfo{journal}{Opt. Lett.}} \textbf{\bibinfo{volume}{22}},
  \bibinfo{pages}{301--303} (\bibinfo{year}{1997}).
\newblock \urlprefix\url{http://ol.osa.org/abstract.cfm?URI=ol-22-5-301}.

\bibitem{Jones28042000}
\bibinfo{author}{Jones, D.~J.} \emph{et~al.}
\newblock \bibinfo{title}{Carrier-envelope phase control of femtosecond
  mode-locked lasers and direct optical frequency synthesis}.
\newblock \emph{\bibinfo{journal}{Science}} \textbf{\bibinfo{volume}{288}},
  \bibinfo{pages}{635--639} (\bibinfo{year}{2000}).
\newblock
  \urlprefix\url{http://www.sciencemag.org/content/288/5466/635.abstract}.

\bibitem{PhysRevLett.82.3568}
\bibinfo{author}{Udem, T.}, \bibinfo{author}{Reichert, J.},
  \bibinfo{author}{Holzwarth, R.} \& \bibinfo{author}{H\"ansch, T.~W.}
\newblock \bibinfo{title}{{Absolute optical frequency measurement of the cesium
  ${\mathit{D}}_{1}$ line with a mode-locked laser}}.
\newblock \emph{\bibinfo{journal}{Phys. Rev. Lett.}}
  \textbf{\bibinfo{volume}{82}}, \bibinfo{pages}{3568--3571}
  (\bibinfo{year}{1999}).
\newblock \urlprefix\url{http://link.aps.org/doi/10.1103/PhysRevLett.82.3568}.

\bibitem{Diddams03082001}
\bibinfo{author}{Diddams, S.~A.} \emph{et~al.}
\newblock \bibinfo{title}{{An optical clock based on a single trapped
  $^{199}\mathrm{Hg}^+$ ion}}.
\newblock \emph{\bibinfo{journal}{Science}} \textbf{\bibinfo{volume}{293}},
  \bibinfo{pages}{825--828} (\bibinfo{year}{2001}).
\newblock
  \urlprefix\url{http://www.sciencemag.org/content/293/5531/825.abstract}.

\bibitem{Steinmetz05092008}
\bibinfo{author}{Steinmetz, T.} \emph{et~al.}
\newblock \bibinfo{title}{Laser frequency combs for astronomical observations}.
\newblock \emph{\bibinfo{journal}{Science}} \textbf{\bibinfo{volume}{321}},
  \bibinfo{pages}{1335--1337} (\bibinfo{year}{2008}).
\newblock
  \urlprefix\url{http://www.sciencemag.org/content/321/5894/1335.abstract}.
\newblock \eprint{http://www.sciencemag.org/content/321/5894/1335.full.pdf}.

\bibitem{Nature.421.611}
\bibinfo{author}{Baltu\v{s}ka, A.} \emph{et~al.}
\newblock \bibinfo{title}{Attosecond control of electronic processes by intense
  light fields}.
\newblock \emph{\bibinfo{journal}{Nature}} \textbf{\bibinfo{volume}{421}},
  \bibinfo{pages}{611--615} (\bibinfo{year}{2003}).
\newblock \urlprefix\url{http://dx.doi.org/10.1038/nature01414}.

\bibitem{PhysRevLett.96.153001}
\bibinfo{author}{Stowe, M.~C.}, \bibinfo{author}{Cruz, F.~C.},
  \bibinfo{author}{Marian, A.} \& \bibinfo{author}{Ye, J.}
\newblock \bibinfo{title}{High resolution atomic coherent control via spectral
  phase manipulation of an optical frequency comb}.
\newblock \emph{\bibinfo{journal}{Phys. Rev. Lett.}}
  \textbf{\bibinfo{volume}{96}}, \bibinfo{pages}{153001}
  (\bibinfo{year}{2006}).
\newblock
  \urlprefix\url{http://link.aps.org/doi/10.1103/PhysRevLett.96.153001}.

\bibitem{Kohler2012159}
\bibinfo{author}{Kohler, M.~C.}, \bibinfo{author}{Pfeifer, T.},
  \bibinfo{author}{Hatsagortsyan, K.~Z.} \& \bibinfo{author}{Keitel, C.~H.}
\newblock \bibinfo{title}{{Chapter 4 - Frontiers of atomic high-harmonic
  generation}}.
\newblock In \bibinfo{editor}{Berman, P.}, \bibinfo{editor}{Arimondo, E.} \&
  \bibinfo{editor}{Lin, C.} (eds.) \emph{\bibinfo{booktitle}{Advances in
  Atomic, Molecular, and Optical Physics}}, vol.~\bibinfo{volume}{61},
  \bibinfo{pages}{159 -- 208} (\bibinfo{publisher}{Academic Press},
  \bibinfo{address}{New York}, \bibinfo{year}{2012}).
\newblock
  \urlprefix\url{http://www.sciencedirect.com/science/article/pii/B9780123964823000041}.

\bibitem{RoehlsbergNature}
\bibinfo{author}{R\"ohlsberger, R.}, \bibinfo{author}{Wille, H.-C.},
  \bibinfo{author}{Schlage, K.} \& \bibinfo{author}{Sahoo, B.}
\newblock \bibinfo{title}{Electromagnetically induced transparency with
  resonant nuclei in a cavity}.
\newblock \emph{\bibinfo{journal}{Nature}} \textbf{\bibinfo{volume}{482}},
  \bibinfo{pages}{199--203} (\bibinfo{year}{2012}).
\newblock \urlprefix\url{http://dx.doi.org/10.1038/nature10741}.

\bibitem{PhysRevA.88.063402}
\bibinfo{author}{Cavaletto, S.~M.}, \bibinfo{author}{Harman, Z.},
  \bibinfo{author}{Buth, C.} \& \bibinfo{author}{Keitel, C.~H.}
\newblock \bibinfo{title}{X-ray frequency combs from optically controlled
  resonance fluorescence}.
\newblock \emph{\bibinfo{journal}{Phys. Rev. A}} \textbf{\bibinfo{volume}{88}},
  \bibinfo{pages}{063402} (\bibinfo{year}{2013}).
\newblock \urlprefix\url{http://link.aps.org/doi/10.1103/PhysRevA.88.063402}.

\bibitem{Zuoye}
\bibinfo{author}{Liu, Z.} \emph{et~al.}
\newblock \bibinfo{title}{Generation of high-frequency combs locked to atomic
  resonances by quantum phase modulation}.
\newblock \emph{\bibinfo{journal}{arXiv:1309.6335}}  (\bibinfo{year}{2013}).
\newblock \urlprefix\url{http://arxiv.org/abs/1309.6335}.

\bibitem{Ott10052013}
\bibinfo{author}{Ott, C.} \emph{et~al.}
\newblock \bibinfo{title}{{Lorentz meets Fano in spectral line shapes: A
  universal phase and its laser control}}.
\newblock \emph{\bibinfo{journal}{Science}} \textbf{\bibinfo{volume}{340}},
  \bibinfo{pages}{716--720} (\bibinfo{year}{2013}).
\newblock
  \urlprefix\url{http://www.sciencemag.org/content/340/6133/716.abstract}.

\bibitem{doi:10.1146/annurev.pc.43.100192.002433}
\bibinfo{author}{Pollard, W.~T.} \& \bibinfo{author}{Mathies, R.~A.}
\newblock \bibinfo{title}{Analysis of femtosecond dynamic absorption spectra of
  nonstationary states}.
\newblock \emph{\bibinfo{journal}{Annual Review of Physical Chemistry}}
  \textbf{\bibinfo{volume}{43}}, \bibinfo{pages}{497--523}
  (\bibinfo{year}{1992}).
\newblock
  \urlprefix\url{http://www.annualreviews.org/doi/abs/10.1146/annurev.pc.43.100192.002433}.
\newblock \bibinfo{note}{PMID: 1463575}.

\bibitem{Marian17122004}
\bibinfo{author}{Marian, A.}, \bibinfo{author}{Stowe, M.~C.},
  \bibinfo{author}{Lawall, J.~R.}, \bibinfo{author}{Felinto, D.} \&
  \bibinfo{author}{Ye, J.}
\newblock \bibinfo{title}{United time-frequency spectroscopy for dynamics and
  global structure}.
\newblock \emph{\bibinfo{journal}{Science}} \textbf{\bibinfo{volume}{306}},
  \bibinfo{pages}{2063--2068} (\bibinfo{year}{2004}).
\newblock
  \urlprefix\url{http://www.sciencemag.org/content/306/5704/2063.abstract}.

\bibitem{Scully:QuantumOptics}
\bibinfo{author}{Scully, M.~O.} \& \bibinfo{author}{Zubairy, S.}
\newblock \emph{\bibinfo{title}{Quantum Optics}} (\bibinfo{publisher}{Cambridge
  University Press}, \bibinfo{address}{Cambridge}, \bibinfo{year}{1997}).

\bibitem{PhysRevA.81.022507}
\bibinfo{author}{Yerokhin, V.~A.} \& \bibinfo{author}{Pachucki, K.}
\newblock \bibinfo{title}{Theoretical energies of low-lying states of light
  helium-like ions}.
\newblock \emph{\bibinfo{journal}{Phys. Rev. A}} \textbf{\bibinfo{volume}{81}},
  \bibinfo{pages}{022507} (\bibinfo{year}{2010}).
\newblock \urlprefix\url{http://link.aps.org/doi/10.1103/PhysRevA.81.022507}.

\bibitem{PhysRevA.56.1288}
\bibinfo{author}{Derevianko, A.} \& \bibinfo{author}{Johnson, W.~R.}
\newblock \bibinfo{title}{{Two-photon decay of $2{}^{1}{S}_{0}$ and
  $2{}^{3}{S}_{1}$ states of heliumlike ions}}.
\newblock \emph{\bibinfo{journal}{Phys. Rev. A}} \textbf{\bibinfo{volume}{56}},
  \bibinfo{pages}{1288--1294} (\bibinfo{year}{1997}).
\newblock \urlprefix\url{http://link.aps.org/doi/10.1103/PhysRevA.56.1288}.

\bibitem{DBLP:journals/cphysics/JonssonHFG07}
\bibinfo{author}{J{\"o}nsson, P.}, \bibinfo{author}{He, X.},
  \bibinfo{author}{Froese~Fischer, C.} \& \bibinfo{author}{Grant, I.~P.}
\newblock \bibinfo{title}{{The \texttt{grasp2K} relativistic atomic structure
  package}}.
\newblock \emph{\bibinfo{journal}{Comput. Phys. Commun.}}
  \textbf{\bibinfo{volume}{177}}, \bibinfo{pages}{597--622}
  (\bibinfo{year}{2007}).

\end{thebibliography}

\end{document}

\title{Supplemental information}

\author{Stefano M. Cavaletto}
\email[Corresponding author. Email: ]{smcavaletto@gmail.com}
\affiliation{Max-Planck-Institut f\"{u}r Kernphysik, Saupfercheckweg~1, 69117~Heidelberg, Germany}
\author{Zolt\'an Harman}
\affiliation{Max-Planck-Institut f\"{u}r Kernphysik, Saupfercheckweg~1, 69117~Heidelberg, Germany}
\author{Christian Ott}
\affiliation{Max-Planck-Institut f\"{u}r Kernphysik, Saupfercheckweg~1, 69117~Heidelberg, Germany}
\author{Christian Buth}
\altaffiliation[Present address: ]{Max-Planck-Institut f\"ur Quantenoptik, Hans-Kopfermann-Stra\ss{}e~1, 85748~Garching bei M\"unchen, Germany}
\affiliation{Max-Planck-Institut f\"{u}r Kernphysik, Saupfercheckweg~1, 69117~Heidelberg, Germany}
\author{Thomas Pfeifer}
\affiliation{Max-Planck-Institut f\"{u}r Kernphysik, Saupfercheckweg~1, 69117~Heidelberg, Germany}
\author{Christoph H. Keitel}
\affiliation{Max-Planck-Institut f\"{u}r Kernphysik, Saupfercheckweg~1, 69117~Heidelberg, Germany}

%letteraaa

\maketitle
 
\section{Equations of motion}
In the article, we assume that the optical transition $2\leftrightarrow 3$ in the three-level system from Fig.~1 is driven by an optical laser field $\boldsymbol{\mathcal{E}}_{\mathrm{L}}(t) = \bar{\mathcal{E}}_{\mathrm{L}}(t)\,\cos(\omega_{\mathrm{L}}t)\,\hat{\boldsymbol{e}}_{\mathrm{L}}$, with envelope $\bar{\mathcal{E}}_{\mathrm{L}}(t)$, central frequency $\omega_{\mathrm{L}}$, and polarization vector $\hat{\bs{e}}_{\mathrm{L}}$. The dynamical evolution of the coherences $\rho_{12}(t)$ and $\rho_{13}(t)$ upon interaction with an optical field $\boldsymbol{\mathcal{E}}_{\mathrm{L}}(t)$ is the solution of the following set of coupled, linear ordinary differential equations\cite{Scully:QuantumOptics}:
\begin{subequations}
\begin{align}
\frac{\diff \rho_{12}}{\diff t} &=  \Bigl(\uimm\,\omega_{21} - \frac{\varGamma_{2}}{2} \Bigr)\rho_{12}(t) + \uimm \frac{\varOmega_{\mathrm{RL}}(t)}{2}\,\eu^{\uimm \omega_{\mathrm{L}}t}\,\rho_{13}(t),\\
\frac{\diff \rho_{13}}{\diff t} &= \Bigl(\uimm\, \omega_{31} - \frac{\varGamma_{3}}{2}\Bigr) \rho_{13}(t) + \uimm \frac{\varOmega_{\mathrm{RL}}(t)}{2}\,\eu^{-\uimm\omega_{\mathrm{L}}t}\,\rho_{12}(t).
\end{align}
\label{eq:firstorderequation}
\end{subequations}
Here, $\omega_{21}$ and $\omega_{31}$ are the transition energies between the ground state and the two excited levels 2 or 3, respectively, while $\varGamma_2$ and $\varGamma_3$ are the total decay rates of the two excited levels. Furthermore, $\varOmega_{\mathrm{RL}}(t)= \bs{d}_{23}\cdot \hat{\bs{e}}_{\mathrm{L}}\,\bar{\mathcal{E}}_{\mathrm{L}}(t)$ is the instantaneous Rabi frequency, assumed to be a real function, provided that the electric field is polarized along the polarization direction of the electric dipole-moment matrix element $\bs{d}_{23}$ of the $2\leftrightarrow 3$ transition. We introduce the slowly oscillating variables
\begin{subequations}
\begin{align}
\bar{\rho}_{12}(t) &= \rho_{12}(t)\,\eu^{-\uimm\omega_{21}(t - T_0)}\,\eu^{\frac{\varGamma_2}{2}(t - T_0)},\\
\bar{\rho}_{13}(t) &= \rho_{13}(t)\,\eu^{-\uimm\omega_{31}(t - T_0)}\,\eu^{\frac{\varGamma_3}{2}(t - T_0)},
\end{align}
\end{subequations}
with respect to the initial time $T_0$. From Eq.~(\ref{eq:firstorderequation}), these variables satisfy the following set of coupled differential equations,
\begin{subequations}
\begin{align}
\frac{\diff \bar{\rho}_{12}}{\diff t} &=  \uimm \frac{\varOmega_{\mathrm{RL}}}{2}\bar{\rho}_{13}(t) \,\eu^{\left(\uimm \varDelta_{\mathrm{L}}  +\frac{\varGamma_{2} - \varGamma_{3}}{2}\right) (t-T_0)},\\
\frac{\diff \bar{\rho}_{13}}{\diff t} &= \uimm \frac{\varOmega_{\mathrm{RL}}}{2}\bar{\rho}_{12}(t)\,\eu^{\left(-\uimm \varDelta_{\mathrm{L}}  - \frac{\varGamma_{2} - \varGamma_{3}}{2}\right) (t-T_0)},
\end{align}
\end{subequations}
with the detuning $\varDelta_{\mathrm{L}} = \omega_{\mathrm{L}} - (\omega_{21} - \omega_{31})$. These two differential equations combined lead to the following second-order differential equation:
\begin{equation}
\begin{split}
\frac{\diff^2 \bar{\rho}_{12}}{\diff t^2} - \left( \frac{1}{\varOmega_{\mathrm{RL}}}\,\frac{\diff \varOmega_{\mathrm{RL}}}{\diff t}  + \uimm \varDelta_{\mathrm{L}} + \frac{\varGamma_{2} - \varGamma_{3}}{2}\right) \frac{\diff \bar{\rho}_{12}}{\diff t} \\
+ \frac{\varOmega_{\mathrm{RL}}^2}{4}\bar{\rho}_{12} =0.
\end{split}
\label{eq:secondorderequation}
\end{equation}
Finally, we define the pulse area 
\begin{equation}
Q_{\mathrm{L}} =\int_{-\infty}^{\infty}\varOmega_{\mathrm{RL}}(t)\,\diff t.
\label{eq:Qinit}
\end{equation}

The previous definitions, valid for a general optical field interacting with the three-level system of Fig.~1 in the article, will be used in the following to study the dynamics of the system in the presence of the two fields,  $\boldsymbol{\mathcal{E}}_{\mathrm{L1}}(t)$ and $\boldsymbol{\mathcal{E}}_{\mathrm{L2}}(t)$, respectively displayed in Fig.1(a) and 1(b) in the main text.%, with corresponding envelopes, $\bar{\mathcal{E}}_{\mathrm{L1}}(t)$ and $\bar{\mathcal{E}}_{\mathrm{L2}}(t)$, and central frequencies, $\omega_{\mathrm{L1}}$ and $\omega_{\mathrm{L2}}$.  

\section{Preparation of the system}

The system is prepared in its initial state via the x-ray pulse X1 and the immediately following optical laser pulse L1 [see Figs.~1(a) and 1(c) in the main text]. The optical field consists of a single pulse $\bs{\mathcal{E}}_{\mathrm{L1}}(t)$ with pulse area $Q_{\mathrm{L1}} = \pi$ and is tuned to the corresponding optical transition with energy $\omega_{23}$. The state of the system at time $T_0$ after the interaction with the \mbox{x-ray} pulse X1 is given by $\rho_{12}(T_0) = \bar{\rho}_{12}(T_0) = \uimm S$, $\rho_{13}(T_0) = \bar{\rho}_{13}(T_0) = 0$. The value of $S$, i.e., the coherence which results from the interaction with the \mbox{x-ray} driving field, depends on the actual envelope and phase of the \mbox{x-ray} pulse, e.g., from a free-electron laser (FEL). { The interaction of the system with the driving pulse X1 can be investigated explicitly by solving the master equation for all the elements of the $3\times 3$ density matrix $\hat{\rho}$.} We do not show this here and assume for simplicity that $S = 1/2$. This condition corresponds to the case in which the \mbox{x-ray} pulse is sufficiently long to transfer half of the population, initially in the ground state 1, into the excited state 2. Such condition could be achieved with the presently available pulses from a FEL. The essence of our results is not affected by the value of $S$, which is only restricted to satisfy the following inequalities, $0\leq|S|\leq1/2$, to ensure the Hermiticity and positive definiteness of the density matrix. We notice that $\rho_{13}(T_0)$ vanishes because the transition between states 1 and 3 is not allowed by single-photon processes and thus cannot be driven by the FEL \mbox{x-ray} pulse. 

With these initial conditions, the second-order differential equation~(\ref{eq:secondorderequation}) can be solved analytically, if the envelope of the driving field $\boldsymbol{\mathcal{E}}_{\mathrm{L1}}(t)$ is modelled via a hyperbolic-secant function \cite{abramowitz1964handbook},
\begin{equation}
\bar{\mathcal{E}}_{\mathrm{L1}}(t) = \bar{\mathcal{E}}_{\mathrm{L1,max}}\, \sech{[\gamma_{\mathrm{1}} (t - t_0)]}.
\label{eq:singlepulse}
\end{equation}
Here, $\bar{\mathcal{E}}_{\mathrm{L1,max}} = \sqrt{8\pi\alpha I_{\mathrm{L1}}}$ is the maximum of the envelope, $I_{\mathrm{L1}}$ is the peak intensity, $\sech(x) = 1/\cosh{(x)}$ is the hyperbolic-secant function, with pulse width $\gamma_{\mathrm{1}}$, corresponding to a FWHM duration of $|\bar{\mathcal{E}}_{\mathrm{L1}}(t)|^2$ given by $\tau_{\mathrm{1}} = 2\arccosh{(\sqrt{2})}/\gamma_{\mathrm{1}}$. The linear polarization vector, the central frequency, and the detuning from the optical transition of the optical pulse L1 are given by $\hat{\bs{e}}_{\mathrm{L1}}$, $\omega_{\mathrm{L1}}$, and $\varDelta_{\mathrm{L1}}$, respectively. The instantaneous Rabi frequency follows to
\begin{equation}
\varOmega_{\mathrm{RL1}}(t) = A_{\mathrm{1}} \sech{[\gamma_{\mathrm{1}} (t - t_0)]},
\label{eq:varOmegagg}
\end{equation}
where $A_{\mathrm{1}} = \bs{d}_{23}\cdot \hat{\bs{e}}_{\mathrm{L1}}\,\bar{\mathcal{E}}_{\mathrm{L1,max}}$ is the peak Rabi frequency\cite{Scully:QuantumOptics}. For a single hyperbolic-secant pulse, the pulse area  follows with Eqs.~(\ref{eq:Qinit}) and (\ref{eq:varOmegagg}) to
\begin{equation}
Q_{\mathrm{L1}} = A_{\mathrm{1}} \int_{-\infty}^{\infty} \sech{[\gamma_{\mathrm{1}} (t - t_0)]} \,\diff t = \pi A_{\mathrm{1}}/\gamma_{\mathrm{1}}.
\end{equation}

We assume that $\gamma_{\mathrm{1}}\gg \varGamma_2\gg \varGamma_3${ , i.e., in particular, the optical pulse L1 is shorter than the decay time of the system}. Furthermore, we also assume that the time delay between the pulses L1 and X1, given by $t_0 - T_0$, is larger than the duration of the optical pulse L1, such that there is no temporal overlap between the two pulses [see Fig.~1(a)]. { By requiring that $t_0 - T_0\ll 1/\varGamma_2$, we also ensure that at the arrival of the pulse L1 only a small fraction of the initial coherence $\rho_{12}(T_0)$ has deteriorated as a result of spontaneous decay.} Under the just described conditions, %i.e., $\varOmega_{\mathrm{RL1}} = A \sech{[\gamma (t - t_0)]}$ with $A = \gamma$, $a = 1/2$, and $c = 1/2 - (\varGamma_{2} - \varGamma_{3})/(4\gamma)$, 
the solution of the equations of motion (EOMs)~(\ref{eq:firstorderequation}) becomes\cite{abramowitz1964handbook}
% \begin{subequations}
% \begin{equation}
% \bar{\rho}_{12}(t) = \uimm \,S \,_2F_1[a,\,-a;\,c\,;z(t,\,t_0)] ,
% \end{equation}
% \begin{equation}
% \bar{\rho}_{13}(z) = -S\,\frac{a}{c}\,\eu^{2\gamma\left(c-\frac{1}{2}\right) (t_0 - T_0)}\, z(t,\,t_0)^c [1-z(t,\,t_0)]^{1-c} \,_2F_1[1-a,\,1+a;\,1+c;\,z(t,\,t_0)],
% \end{equation}
% \end{subequations}
% and
\begin{subequations}
\begin{equation}
{\rho}_{12}(t) = \uimm\,S\,\eu^{\uimm\omega_{21}(t-T_0)}\,\eu^{- \frac{\varGamma_2}{2}(t-T_0)}\,_2F_1\left[a,\,-a;\,c;\,z(t,\,t_0)\right],
\end{equation}
\begin{equation}
\begin{split}
{\rho}_{13}(t) =&\, -S\,\frac{a}{c}\,\eu^{- \frac{\varGamma_2}{2}(t_0-T_0)}\,\eu^{\uimm(\omega_{31} - \varDelta_{\mathrm{L1}})(t-T_0)}\\
&\times\,\eu^{\left(\uimm\varDelta_{\mathrm{L1}}-\frac{\varGamma_3}{2}\right)(t-t_0)}\,[z(t,\,t_0)]^c\, [1-z(t,\,t_0)]^{1-c}\\
&\times \,_2F_1[1-a,\,1+a;\,1+c;\,z(t,\,t_0)].
\end{split}
\end{equation}
\label{eq:thesolutionsinglepulse}
\end{subequations}
where $\,_2F_1(a,\,b;\,c;\,z)$ is the Gaussian (or ordinary) hypergeometric function\cite{abramowitz1964handbook} and where we have defined the two parameters
\begin{subequations}
\begin{align}
a &= \frac{A_{\mathrm{1}}}{2\gamma_{\mathrm{1}}},\\
c &= \frac{1}{2}- \uimm \frac{\varDelta_{\mathrm{L1}}}{2\gamma_{\mathrm{1}}} - \frac{\varGamma_{2} - \varGamma_{3}}{4\gamma_{\mathrm{1}}},
\end{align}
\end{subequations}
and the function
\begin{equation}
z(t,\,t_0) = \frac{\tanh{[\gamma_{\mathrm{1}} (t-t_0)]} + 1}{2}.
\end{equation}

\begin{figure}[tb]
\centering%
\includegraphics[width=0.75\linewidth, keepaspectratio]{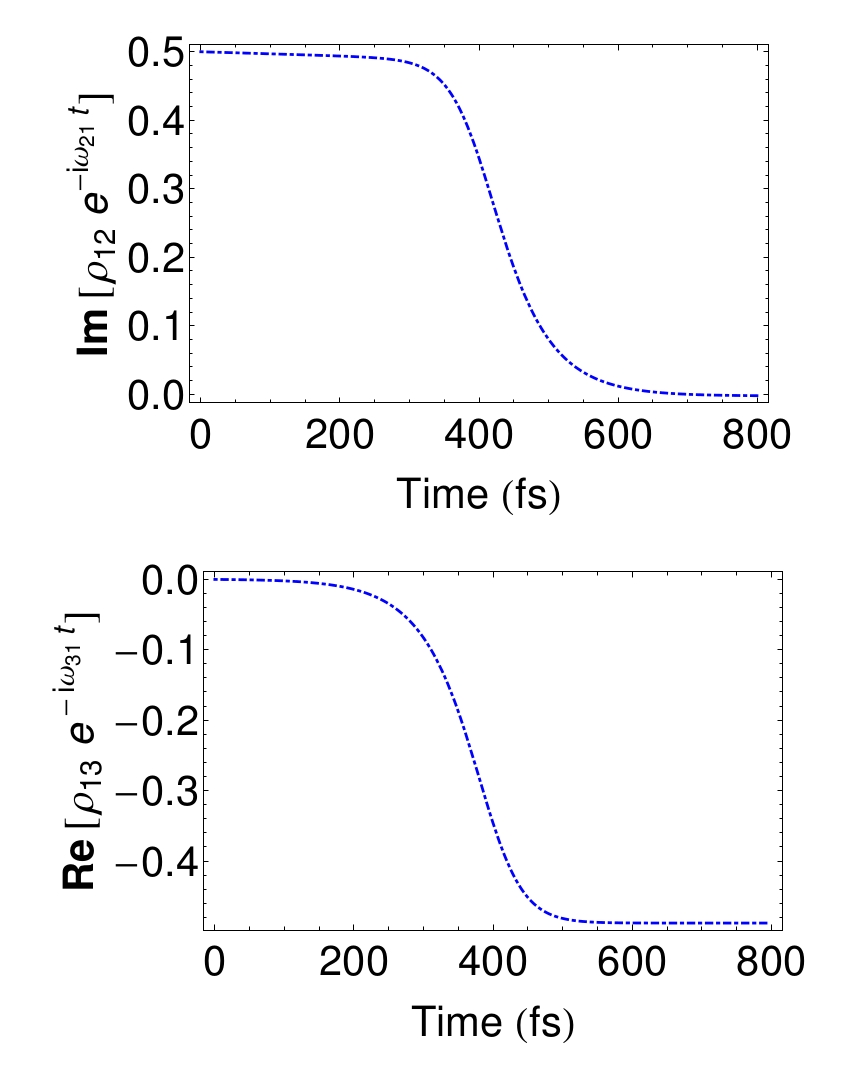}
\caption{\textbf{Time evolution of the atomic system employing a single optical pulse}. The $\Lambda$-type, three-level system introduced in the main text is applied to heliumlike $\mathrm{Be}^{2+}$ ions. The transition energies\cite{PhysRevA.81.022507} are $\omega_{21} = 123.7\unit{eV}$ and $\omega_{23}= 2.02\,\mathrm{eV}$, while the decay rate is given by $\varGamma_2 =1.2\times 10^{11}\unit{s^{-1}} $, computed with \texttt{grasp2K}\cite{DBLP:journals/cphysics/JonssonHFG07}. The two-photon decay rate $\varGamma_3$ is comparably negligible. The $2\leftrightarrow 3$ transition is driven by a single pulse L1 with envelope~(\ref{eq:varOmegagg}), tuned to the optical transition. We plot $\Imm[{\rho}_{12}(t)\,\eu^{-\uimm\omega_{21}t}]$ (top panel) and $\Real[{\rho}_{13}(t)\,\eu^{-\uimm\omega_{21}t}]$ (bottom panel), by assuming an initial state at $T_0 = 0$ given by $\rho_{12}(T_0) = \uimm/2$, $\rho_{13}(T_0) = 0$ resulting from the excitation with the \mbox{x-ray} pulse X1. The pulse L1 has a FWHM duration of $\tau_1= 100\,\unit{fs}$, corresponding to a spectral width of $\gamma_1 = 0.07\unit{eV}$, and has its maximum at $t_{0} = 400\unit{fs}$, with the peak intensity $I_{\mathrm{L1,max}} = 6.42 \times 10^{9}\unit{W/cm^2}$.}%
\label{fig:ShowSupp}
\end{figure}

%which we plot in Fig.\ref{Fig:} for different values of the pulse area (\ref{eq:pulsearea}) $Q = \frac{A}{\gamma} \pi = 2 a \pi$ and for $\bar{\varrho}_{12,\mathrm{in}} = \frac{1}{2}$.
By defining
\begin{subequations}
\begin{equation}
\mu_a = \lim_{t\rightarrow \infty} \,_2F_1\left(a,-a;c;z(t,\,t_0)\right) = \frac{[\Gamma(c)]^2}{\Gamma(c-a)\,\Gamma(c+a)}, %\ \text{and, if $\Real{(c-a)}=0$, exactly equal to 0},
\end{equation}
\begin{equation}
\begin{split}
\nu_a =&\, \lim_{t\rightarrow \infty} [z(t,\,t_0)]^{c}  \\
&\,\times\,[1-z(t,\,t_0)]^{1-c} \,_2F_1[1-a,\,1+a;\,1+c;\,z(t,\,t_0)]\\
=&\, \frac{\Gamma(1+c)\,\Gamma(1-c)}{\Gamma(1-a)\,\Gamma(1+a)} = \frac{c\,\csc{(\pi c)}}{a\,\csc{(\pi a)}},
\end{split}
\end{equation}
\label{eq:finalvaluespulsemunu}
\end{subequations}
with $\csc(x) = 1/\sin(x)$ and $\Gamma(x)$ being the cosecant and the Gamma function, respectively, we can see that, in the long-time limit, the two off-diagonal matrix elements of the density operator $\rho_{12}(t)$ and $\rho_{13}(t)$ behave like
\begin{subequations}
\begin{equation}
\lim_{t\rightarrow \infty}{\rho}_{12}(t)= \uimm\,S\,\mu_a\,\eu^{\uimm\omega_{21}(t-T_0)}\,\eu^{- \frac{\varGamma_2}{2}(t-T_0)},
\label{eq:limmua}
\end{equation}
\begin{equation}
\begin{split}
\lim_{t\rightarrow \infty}{\rho}_{13}(t) = &\,-S\,\frac{a\,\nu_a}{c}\,\eu^{- \left(\uimm\varDelta_{\mathrm{L1}}+\frac{\varGamma_2}{2}\right)(t_0-T_0)}\,\eu^{\uimm\omega_{31}(t-T_0)}\\
&\,\times\,\eu^{-\frac{\varGamma_3}{2}(t-t_0)}.
\label{eq:limnua}
\end{split}
\end{equation}
\end{subequations}
The aforementioned limit $t\rightarrow\infty$ is already realized right after the single pulse L1. The two parameters, therefore, describe the state of the system immediately after that the interaction with the optical pulse has concluded. In particular, $\mu_a$ allows one to quantify the effect of the pulse on the variable $\rho_{12}(t)$. For times following the interaction with the optical pulse, the function $\rho_{12}(t)$ in Eq.~(\ref{eq:limmua}) freely decays with decay rate $\varGamma_2$ from a new, effective initial value $\uimm\,S\,\mu_a$. The behaviour of the coherence $\rho_{13}(t)$ in Eq.~(\ref{eq:limnua}) after the interaction with the optical pulse can be similarly explained, i.e., the function decays with the much smaller decay rate $\varGamma_3$ from a new, effective initial value $-S\,(a\,\nu_a/c)\,\eu^{- \left(\uimm\varDelta_{\mathrm{L1}}+\frac{\varGamma_2}{2}\right)(t_0-T_0)}$.

Because of the poles of the Gamma function $\Gamma(x)$ for negative integers\cite{abramowitz1964handbook} $x = -n$, we observe that $\mu_a = 0$ if $c = a$, i.e., if
\begin{subequations}
\begin{align}
\varDelta_{\mathrm{L1}} &= 0,\\
\frac{Q_{\mathrm{L1}}}{2\pi} & = \frac{1}{2} - \frac{\varGamma_{2} - \varGamma_{3}}{4\gamma_{\mathrm{1}}} \approx \frac{1}{2}.
\end{align}
\label{eq:condizioni}
\end{subequations}
A vanishing value of $\mu_a$ guarantees that, in the long-time limit, the off-diagonal matrix element $\rho_{12}(t)$ in Eq.~(\ref{eq:limmua}) is also equal to 0. 

The preparatory stage of which we take advantage in Fig.~1(a) in the main text exploits Eq.~(\ref{eq:condizioni}) to ensure that the system, after having interacted with the \mbox{x-ray} pulse X1 and the optical pulse L1, is prepared in an initial state in which all the coherence is ``stored'' in the slowly decaying function $\rho_{13}(t)$, while $\rho_{12}$ is equal to 0. { With the aforementioned requirements on the duration of the optical pulse and the time delay $t_0 - T_0$, we ensure that spontaneous decay of the excited level 2 does not compromise this mechanism.} As apparent from Eq.~(\ref{eq:condizioni}), this requires the use of an optical pulse L1 which is tuned to the optical transition, $\varDelta_{\mathrm{L1}} = 0$, and with a pulse area of $Q_{\mathrm{L1}} = \pi - \frac{\pi(\varGamma_{2} - \varGamma_{3})}{2\gamma_{\mathrm{1}}} \approx \pi $, such that $\mu_a =0$ and therefore ${\rho}_{12} = 0$. Since the decay rates $\varGamma_{2}$ and $\varGamma_3$ are by orders of magnitude smaller than the spectral width of the optical pulse $\gamma_{\mathrm{1}}$, the area of the pulse is approximately equal to $\pi$. The corresponding evolution of the elements of the density matrix is displayed in Fig.~\ref{fig:ShowSupp}. The effect of a tuned, $\pi$-area pulse is to transfer the coherence from $\rho_{12}(t)$ to $\rho_{13}(t)$. Note that, even if the condition~(\ref{eq:condizioni}) on the pulse area is not exactly met, the rapid decay of $\rho_{12}(t)$, with rate $\varGamma_2$, after the optical pulse L1 ensures that, in a time interval on the order of the repetition period $T_{\mathrm{p}}$ of the subsequent train of optical pulses L2, the residual coherence $\rho_{12}$ between states 1 and 2 will have completely decayed.

\section{Coherence transfer with an optical frequency comb}

\subsection{Interaction with the first pulse}
Let us assume for the time being that the three-level system in Fig.~1(b) in the main text interacts with a single pulse L2. We also assume that the interaction with the \mbox{x-ray} pulse X1 and the optical pulse L1 [Fig.~1(a) in the main text] has prepared the system in an initial state which, owing to Eq.~(\ref{eq:condizioni}), is given by $\rho_{12}(T_0) = \bar{\rho}_{12}(T_0) =0$, $\rho_{13}(T_0) = \bar{\rho}_{13}(T_0) = - V$. The initial time $T_0$ now shall follow the whole preparatory step, i.e., after both the \mbox{x-ray} pulse X1 and the optical pulse L1. Furthermore, $V$ is the value of the off-diagonal matrix element $\rho_{13}$ obtained when the interaction with the optical pulse L1 has concluded. The subsequent optical pulse L2 is centred at $t_0$, such that there is no temporal overlap between this pulse and the (FEL \mbox{x-ray} and optical) pulses adopted to prepare the system in its initial state. The pulse L2 has central frequency $\omega_{\mathrm{L2}}$, polarization $\hat{\bs{e}}_{\mathrm{L2}}$, and detuning $\varDelta_{\mathrm{L2}} = \omega_{\mathrm{L2}} - (\omega_{21} - \omega_{31})$. In order to solve the differential equation~(\ref{eq:secondorderequation}) analytically, we use again a pulse with a hyperbolic-secant envelope,
\begin{equation}
\bar{\mathcal{E}}_{\mathrm{L2}}(t) = \bar{\mathcal{E}}_{\mathrm{L2,max}} \sech{\left[\gamma_2\left(t - t_0\right)\right]}.
\end{equation}
The instantaneous Rabi frequency is
\begin{equation}
\varOmega_{\mathrm{RL2}}(t) = A_{\mathrm{2}} \sech{[\gamma_{\mathrm{2}} (t - t_0)]},
\end{equation}
with $A_{\mathrm{2}} = \bs{d}_{23}\cdot \hat{\bs{e}}_{\mathrm{L2}}\,\bar{\mathcal{E}}_{\mathrm{L2,max}}$, and the pulse area is given by
\begin{equation}
Q_{\mathrm{L2}} = \pi A_{\mathrm{2}}/\gamma_{\mathrm{2}}.
\end{equation}
{ We assume that $\gamma_{\mathrm{2}}\gg \varGamma_2\gg \varGamma_3$, i.e., in particular, the optical pulse L2 is shorter than the decay time of the system}. The parameters $a$ and $c$ are now defined as
\begin{subequations}
\begin{align}
a &= \frac{A_{\mathrm{2}}}{2\gamma_{\mathrm{2}}},\\
c &= \frac{1}{2}- \uimm \frac{\varDelta_{\mathrm{L2}}}{2\gamma_{\mathrm{2}}} - \frac{\varGamma_{2} - \varGamma_{3}}{4\gamma_{\mathrm{2}}},
\end{align}
\end{subequations}
while the function $z(t,\,t_0)$ is defined as in the previous section. For the given set of initial conditions, the solution of the EOMs is symmetric to the previously discussed solution~(\ref{eq:thesolutionsinglepulse}), i.e.,
\begin{subequations}
\begin{equation}
\begin{split}
{\rho}_{12}(t) =&\, -{\uimm}\,V\, \frac{a}{1-c}\,\eu^{\uimm\omega_{21} (t-T_0)}\,\eu^{- \frac{\varGamma_3}{2}(t_0-T_0)}\,\eu^{\uimm\varDelta_{\mathrm{L2}}(t_0-T_0)}\\
&\, \times\,\eu^{- \frac{\varGamma_2}{2}(t-t_0)} \,[z(t,\,t_0)]^{1-c}\, [1-z(t,\,t_0)]^{c} \\
&\, \times\,_2F_1[1-a,\,1+a;\,2-c;\,z(t,\,t_0)],
\end{split}
\end{equation}
\begin{equation}
\begin{split}
{\rho}_{13}(t) =&\,-V\,\eu^{\uimm\omega_{31}(t-T_0)}\,\eu^{- \frac{\varGamma_3}{2}(t-T_0)}\\
&\,\times\,_2F_1\left[a,\,-a;\,1-c;\,z(t,\,t_0)\right].
\end{split}
\end{equation}
\label{eq:showthephaseosc1}
\end{subequations}
%which we plot in Fig.\ref{Fig:} for different values of the pulse area (\ref{eq:pulsearea}) $Q = \frac{A}{\gamma} \pi = 2 a \pi$ and for $\bar{\varrho}_{12,\mathrm{in}} = \frac{1}{2}$.
Also in this case we define the two variables
\begin{subequations}
\begin{equation}
\begin{split}
\lambda_a =&\,\lim_{t\rightarrow \infty}[z(t,\,t_0)]^{1-c}\, [1-z(t,\,t_0)]^{c} \\
&\,\times\,_2F_1[1-a,\,1+a;\,2-c;\,z(t,\,t_0)]\\
=&\, \frac{\Gamma(c)\,\Gamma(2-c)}{\Gamma(1-a)\,\Gamma(1+a)}=\frac{(1-c)\,\csc{[\pi (1-c)]}}{a\,\csc{(\pi a)}},
\end{split}
\label{eq:lambdaa}
\end{equation}
\begin{equation}
\begin{split}
\xi_a &\,= \lim_{t\rightarrow \infty} \,_2F_1\left[a,\,-a;\,1-c;\,z(t,\,t_0)\right]\\
&\, = \frac{[\Gamma(1-c)]^2}{\Gamma(1-c-a)\,\Gamma(1-c+a)}, 
\end{split}
%\ \text{and, if $\Real{(c-a)}=0$, exactly equal to 0}, 
\end{equation}
\label{eq:finalvaluespulselambdaxi}
\end{subequations}
which describe the action of the optical pulse L2 on the atomic coherences. In particular, we observe that
\begin{subequations}
\begin{equation}
\begin{split}
\lim_{t\rightarrow \infty}{\rho}_{12}(t)=&\,  -\uimm\,V\,\frac{\lambda_a\,a}{1-c}\,\eu^{\uimm\omega_{21} (t-T_0)}\,\eu^{-\left(\uimm\varDelta_{\mathrm{L2}}+  \frac{\varGamma_3}{2}\right)(t_0-T_0)}\\
&\,\times\,\eu^{- \frac{\varGamma_2}{2}(t-t_0)},
\end{split}
\end{equation}
\begin{equation}
\lim_{t\rightarrow \infty}{\rho}_{13}(t)= -V\,\xi_a\,\eu^{\uimm\omega_{31}(t-T_0)}\,\eu^{- \frac{\varGamma_3}{2}(t-T_0)}.
\end{equation}
\end{subequations}
The aforementioned limit $t\rightarrow\infty$ is already realized right after the pulse L2. The value of $\lambda_a$ vanishes (i) for $a = n$, $n\in\mathbb{N}$, i.e., for a pulse area $Q_{\mathrm{L2}}$ which is an integer multiple of $2\pi$, or (ii) in the large-detuning limit, i.e., when $\varDelta_{\mathrm{L2}} \gg \gamma_2$. This means that, when one of these two conditions is satisfied, the coherence $\rho_{12}(t)$ after interaction with the optical pulse L2 is led back to its vanishing initial value. As a result, the time interval in which $\rho_{12}(t)$ differs from 0 is given by the duration of the pulse itself. Its envelope results in a pulse-shape function which is as short as the envelope of the driving optical pulse. 

The factor $\xi_a$ represents the modification of $\rho_{13}(t)$ in phase and amplitude as a result of the interaction with the optical pulse L2. The function $\rho_{13}(t)$ undergoes a phase shift given by $\arg{(\xi_a)}$ and a decrease in its amplitude given by $|\xi_a|<1$.

\subsection{Interaction with the periodic train of pulses}

The train of pulses from an optical frequency comb $\boldsymbol{\mathcal{E}}_{\mathrm{L2}}(t) = \bar{\mathcal{E}}_{\mathrm{L2}}(t)\,\cos(\omega_{\mathrm{L2}}t )\,\hat{\boldsymbol{e}}_{\mathrm{L2}}$,
which drives the three-level system of Fig.~1(b) in the main text, is described by the carrier frequency $\omega_{\mathrm{L2}}$ and by the periodic envelope $\bar{\mathcal{E}}_{\mathrm{L2}}(t)$, which we model as
\begin{equation}
\bar{\mathcal{E}}_{\mathrm{L2}}(t) = \bar{\mathcal{E}}_{\mathrm{L2,max}}  \sum_{j=0}^{\infty}  \sech{\left[\gamma_2\left(t - t_j\right)\right]}.
\label{eq:envelopEL2}
\end{equation}
The envelope of the field L2 consists of a train of hyperbolic-secant pulses located at $t_j = t_0 + j\, T_{\mathrm{p}}$, $j\in \mathbb{N}_0$, with repetition period $T_{\mathrm{p}}$. The other defining parameters of the comb, such as peak intensity, detuning, FWHM duration, and spectral bandwidth, were already introduced in the preceding section while analysing the results for a single pulse. In this case, the pulse area $Q_{\mathrm{L2}}$ is defined via the integral of the instantaneous Rabi frequency
\begin{equation}
\varOmega_{\mathrm{RL2}}(t) = A_{\mathrm{2}} \sum_{j=0}^{\infty}  \sech{[\gamma_{\mathrm{2}} (t - t_j)]},
\end{equation}
over a single pulse, i.e., 
\begin{equation}
Q_{\mathrm{L2}} =\int_{t_j}^{t_j + T_{\mathrm{p}}}\varOmega_{\mathrm{RL2}}(t)\,\diff t = \pi A_{\mathrm{2}}/\gamma_{\mathrm{2}}.
\label{eq:Q2}
\end{equation}

For the case $\tau_2\ll T_{\mathrm{p}}$, the envelope function is given by a train of independent pulses. From the single-pulse solution~(\ref{eq:showthephaseosc1}) we know that $\rho_{12}(t)$, after interaction with an optical pulse, is led back to its vanishing initial value if the constant $\lambda_a$ in Eq.~(\ref{eq:lambdaa}) is equal to 0. This condition is fulfilled when either (i) the pulse area $Q_{\mathrm{L2}}$ in Eq.~(\ref{eq:Q2}) is an integer multiple of $2\pi$, or (ii) the detuning is much larger than the pulse width, $\varDelta_{\mathrm{L2}} \gg \gamma_{2}$. When one of these two conditions is met, then the solution of the EOMs~(\ref{eq:secondorderequation}) describing the interaction of the system with each one of the pulses in the periodic envelope $\bar{\mathcal{E}}_{\mathrm{L2}}(t)$ can be obtained from the previously described single-pulse solution~(\ref{eq:showthephaseosc1}). For the initial conditions $\rho_{12}(T_0) = \bar{\rho}_{12}(T_0) =0$, $\rho_{13}(T_0) = \bar{\rho}_{13}(T_0) = - V$, resulting from the interaction of the atomic system with the \mbox{x-ray} pulse X1 and the optical pulse L1 [Fig.~1(a) in the article], the solution of the EOMs is given by
\begin{subequations}
\begin{align}
\begin{split}
&\rho_{12}(t) =- \uimm \,V\,\frac{a}{2(1-c)}\,\eu^{\left[\uimm (\omega_{21} + \varDelta_{\mathrm{L2}}) -\frac{\varGamma_{3}}{2}\right] \left(t - T_0\right)}\\
&\times\sum_{j =0}^{\infty}\xi_a^{\,j}\, \sech{[\gamma_{2}(t-t_j)]} \,_2F_1[1-a,1+a;2-c;z(t,\,t_j)],
\end{split}\\
% \end{equation}%\vspace{-0.7cm}
% \begin{equation}
\begin{split}
&\rho_{13}(t) =-V\,  \eu^{\left(\uimm\omega_{31} - \frac{\varGamma_{3}}{2}\right) \left(t-T_0 \right)}\\
&\times\,\biggl[1+ \sum_{j = 0}^{\infty} \xi_a^{\,j}\times \Bigl( \,_2F_1[a,\,-a;\,1-c;\,z(t,\,t_j)]  - 1\Bigr)\biggr].%\\%&
\end{split}
\end{align}
\label{eq:rho12forFourierBis}
\end{subequations}

In the following, the main features of the obtained solution are described. The envelope of $\rho_{12}(t)$ is given by a periodic train of hyperbolic-secant functions, centred at $t_j = t_0 + j T_{\mathrm{p}}$ and with the same FWHM duration as the optical pulses in the optical frequency comb. The phase of $\rho_{12}(t)$ displays fast oscillations given by the term $\eu^{\uimm(\omega_{21}+\varDelta_{\mathrm{L2}})t}$. The amplitudes of both $\rho_{12}(t)$ and $\rho_{13}(t)$ undergo a constant decrease during a period $T_{\mathrm{p}}$, i.e.,
\begin{equation}
\begin{split}
&\,|\rho_{1k}(t + T_{\mathrm{p}})|^2  = \eu^{- \varGamma_3{T_{\mathrm{p}}}}\,|\xi_a|^2\,|\rho_{1k}(t )|^2 \\
= &\,\eu^{- \left(\varGamma_3 - 2\frac{\log(|\xi_a|)}{T_{\mathrm{p}}}\right){T_{\mathrm{p}}}}\,|\rho_{1k}(t )|^2 = \eu^{- \frac{T_{\mathrm{p}}}{\tilde{\tau}_a}}\,|\rho_{1k}(t)|^2 ,
\end{split}
\end{equation}
for $t>t_0$ and $k\in\{2,\,3\}$, which leads to the effective decay time
\begin{equation}
\tilde{\tau}_a = \frac{1}{\varGamma_3 - 2\frac{\log(|\xi_a|)}{T_{\mathrm{p}}}},
\label{eq:tauntilde}
\end{equation} 
describing the decrease of $\rho_{1k}$. This decay time determines the effective decay rate $1/\tilde{\tau}_a$ of the system. 
% It is interesting to notice that the effective decay rate $1/\tilde{\tau}_a$ is in this case only indirectly related to $\varGamma_2$ through the loss of coherence $|\xi_a|$, periodically taking place in the presence of each pulse. In such case, therefore, the coherence is stored in the metastable state $3$, whose decay rate $\varGamma_3$ is negligible if compared to the decay rate of the fast decaying level $2$. The transfer of coherence to state 2, and the connected \mbox{x-ray} photon emission due to spontaneous decay from such level, is limited to the very short time interval in which an optical pulse from the optical frequency comb is present. Finally, we also observe that, at every pulse, an additional phase shift $\arg(\xi_a)$ is caused.

To evaluate the absorption spectrum of the transmitted \mbox{x-ray} pulse X1 in Eq.~(1) in the main text, we perform the Fourier transform
\begin{equation}
\begin{split}
&\int_{T_0}^{\infty}\rho_{12}(t)\,\eu^{-\uimm \omega (t - T_0)}\,\diff t\\
= &\,-{\uimm}V\, \frac{a}{1-c}\,\eu^{\left[- \uimm(\omega -\omega_{21}  - \varDelta_{\mathrm{L2}}) -\frac{\varGamma_3 }{2}\right](t_0-T_0)}\\
&\sum_{m=-\infty}^{\infty}\frac{\eu^{\left[\uimm\left(\omega - \omega_{\mathrm{21}} - \varDelta_{\mathrm{L2}} - \frac{\arg{\xi_a}}{T_{\mathrm{p}}} - \frac{2\pi m}{T_{\mathrm{p}}}\right) + \frac{\varGamma_3}{2} - \frac{\log{|\xi_a|}}{T_{\mathrm{p}}} \right]\frac{T_{\mathrm{p}}}{2}}}{\frac{\varGamma_{3}}{2} - \frac{\log{|\xi_a|}}{T_{\mathrm{p}}} + \uimm \left(\omega - \omega_{21} - \varDelta_{\mathrm{L2}}  - \frac{\arg{\xi_a}}{T_{\mathrm{p}}}  - \frac{2\pi m}{T_{\mathrm{p}}}\right)}\\
& \,\times \, \frac{\pi}{2\gamma_{2} T_{\mathrm{p}}}\, \sech\left[\tfrac{\pi}{2\gamma_{2} T_{\mathrm{p}}} (2 m\pi + \arg{\xi_a} - \uimm\log{|\xi_a|})\right]\\
&\, \times \,_3F_2\bigl[1-a,1+a,\tfrac{1}{2}-\uimm\tfrac{2\pi m+\arg{\xi_a}-\uimm\log{|\xi_a|}}{2\gamma_{2} T_{\mathrm{p}}};2-c, 1; 1\bigr].
\end{split}
\label{eq:thefinalFourier13}
\end{equation}
A comb of equidistant peaks is displayed in Eq.~(\ref{eq:thefinalFourier13}). Each peak has a spectral width given by the effective decay rate $1/\tilde{\tau}_a$ of the atomic variables. Furthermore, the large overall width of the spectrum, i.e., the large number of comb peaks, is a consequence of the short duration of the pulses constituting the envelope of $\rho_{12}(t)$. The $m$th peak is centred at the frequency $\omega_m =  \omega_{\mathrm{21}} + \varDelta_{\mathrm{L2}} + \frac{\arg{\xi_a}}{T_{\mathrm{p}}} + \frac{2\pi m}{T_{\mathrm{p}}}$, with a direct dependence on $\varDelta_{\mathrm{L2}}$ and $\arg(\xi_a)$: This can be understood as a consequence of the oscillating term $\eu^{\uimm(\omega_{21}+\varDelta_{\mathrm{L2}})t}$ and of the periodic phase shift $\arg(\xi_a)$ which characterize the time evolution of the coherence $\rho_{12}(t)$. 

{ \section{Estimation of the emitted power per comb line}

The \mbox{x-ray} pulse-shaping method put forward in the main text uses an optical frequency comb driving an optical transition to rearrange the spectral distribution of the \mbox{x-ray} energy which is absorbed by a sample of ions driven by an initial pulse X1. The total output energy is given by \cite{Foot:AtomicPhysics}
\begin{equation}
E_{\mathrm{X_{out}}} = N L \int \sigma_{\mathrm{abs}}(\omega)\, \frac{\diff E_{\mathrm{X1}}(\omega)}{\diff\omega}\,\diff \omega,
\label{eq:outputxrayen}
\end{equation}
where $N$ is the number density in the atomic sample, $L$ the interaction length, $\sigma_{\mathrm{abs}}(\omega)$ the energy-dependent absorption cross-section, and ${\diff E_{\mathrm{X1}}(\omega)}/{\diff\omega}$ the energy density of the driving pulse tuned to the \mbox{x-ray} transition $\omega_{21}$. For an intensity of $I_{\mathrm{X1}}$, a duration of $\tau_{\mathrm{X1}}$, a beam area of $\mathscr{A}_{\mathrm{X1}}$, and a bandwidth of $\Delta\omega_{\mathrm{X1}}$, we approximate the energy density around the transition energy $\omega_{21}$ with the constant value
\begin{equation}
\frac{\diff E_{\mathrm{X1}}(\omega)}{\diff\omega} = \frac{I_{\mathrm{X1}}\tau_{\mathrm{X1}}\mathscr{A}_{\mathrm{X1}}}{\Delta\omega_{\mathrm{X1}}}.
\end{equation} 

If only the pulse X1 is employed, the resulting absorption cross section \cite{Foot:AtomicPhysics}
\begin{equation}
\sigma^{\text{X1-only}}_{\mathrm{abs}}(\omega)=\frac{3\pi^2 c^2\varGamma_{21}}{\omega_{21}^2}\,\frac{1}{2\pi}\,\frac{\varGamma_{2}}{(\omega - \omega_{21})^2 + \frac{\varGamma_{2}^2}{4}} ,
\end{equation}
with $c$ being the speed of light in vacuum, is given by a Lorentzian function centred on the \mbox{x-ray} transition energy $\omega_{21}$ and with line width equal to the decay rate $\varGamma_2$, such that the following identities are satisfied:
\begin{subequations}
\begin{align}
\int \sigma^{\text{X1-only}}_{\mathrm{abs}}(\omega)\,\diff \omega &= \frac{3 \pi^2 c^2 \varGamma_{21}}{\omega_{21}^2} ,\\
\sigma^{\text{X1-only}}_{\mathrm{abs}}(\omega_{21}) &= \frac{6\pi c^2\varGamma_{21}}{\varGamma_{2}\,\omega_{21}^2}. \label{eq:maxabsence}
\end{align}
\end{subequations}
The corresponding total absorbed energy is given by
\begin{equation}
E^{\text{X1-only}}_{\mathrm{X_{out}}} = N L \frac{3\pi^2 c^2 \varGamma_{21}}{\omega_{21}^2} \frac{I_{\mathrm{X1}}\tau_{\mathrm{X1}}\mathscr{A}_{\mathrm{X1}}}{\Delta\omega_{\mathrm{X1}}}.
\end{equation}

The quantum-control scheme which we present in the manuscript modifies the absorption spectrum [see Eq.~(1) in the main text], by distributing the total energy over a large number of equally separated peaks, centred on the frequencies $\omega_m =  \omega_{\mathrm{21}} + \varDelta_{\mathrm{L2}} + \frac{\arg{\xi_a}}{T_{\mathrm{p}}} + \frac{2\pi m}{T_{\mathrm{p}}}$, with $m$ being an integer number. Because of energy conservation, the width of these peaks decreases correspondingly, being given by the effective decay rate $1/\tilde{\tau}_a$ from Eq.~(\ref{eq:tauntilde}). In Figs.~2 and 3 in the main text we display the optically modified absorption spectrum $\sigma_{\mathrm{abs}}(\omega)$, normalized to the maximum $\sigma^{\text{X1-only}}_{\mathrm{abs}}(\omega_{21})$ of the \mbox{x-ray}-only, single-peak Lorentzian spectrum from Eq.~(\ref{eq:maxabsence}). From the pictures it appears that, at the central frequency $\omega_m$ of the $m$th peak, the ratio $\sigma_{\mathrm{abs}}(\omega_m)/\sigma^{\text{X1-only}}_{\mathrm{abs}}(\omega_{21})$ is of the order of 1. The energy $E_{\mathrm{X_{out}},m}$ in the $m$th peak in the \mbox{x-ray} comb $\mathrm{X_{out}}$ can be estimated by integrating the output \mbox{x-ray} energy~(\ref{eq:outputxrayen}) within an interval of width $1/\tilde{\tau}_a$ around the peak central frequency $\omega_m$, thus obtaining
\begin{equation}
E_{\mathrm{X_{out}},m}\approx N L \frac{\sigma^{\text{X1-only}}_{\mathrm{abs}}(\omega_{21})}{\tilde{\tau}_a} \frac{I_{\mathrm{X1}}\tau_{\mathrm{X1}}\mathscr{A}_{\mathrm{X1}}}{\Delta\omega_{\mathrm{X1}}}.
\end{equation}
We obtain the power per comb peak $P_{\mathrm{X_{out}},m}$ averaged over the effective decay time $\tilde{\tau}_a$ of the atomic system as the energy in the $m$th peak divided by $\tilde{\tau}_a$:
\begin{equation}
P_{\mathrm{X_{out}},m} \approx N L \frac{1}{\tilde{\tau}_a^2}\frac{6\pi c^2\varGamma_{21}}{\varGamma_{2}\,\omega_{21}^2} \frac{I_{\mathrm{X1}}\tau_{\mathrm{X1}}\mathscr{A}_{\mathrm{X1}}}{\Delta\omega_{\mathrm{X1}}}.
\label{eq:Poutm}
\end{equation}

We assume an \mbox{x-ray} pulse with $10^{12}$ photons, centred at the \mbox{x-ray} transition energy $\omega_{21}$, with a duration of $100\,\mathrm{fs}$ and a focal volume with area of $\pi(50\,\mathrm{\mu m})^2$ and length of $2\,\mathrm{cm}$, thus with a peak intensity of $2.5\times10^{12}\,\mathrm{W/cm^2}$, as available at FEL facilities. The bandwidth of the pulse is assumed to be $1\,\mathrm{eV}$. The ensemble of \mbox{Be$^{2+}$} ions which we use to model our quantum-control scheme could be generated, e.g., with an electron-beam ion trap (EBIT), which has already been successfully utilized for highly-charged-ion optical laser spectroscopy\cite{PhysRevLett.107.143002} and \mbox{x-ray} spectroscopy at an \mbox{x-ray} FEL\cite{BernittNature}. We assume therefore an ion density of $10^9\,\mathrm{cm^{-3}}$, as typically reached in an EBIT\cite{BernittNature}. By inserting these parameters in Eq.~(\ref{eq:Poutm}), we predict a power per comb line of $\sim\,30\,\mathrm{pW}$, of the same order of magnitude of \mbox{XUV} combs generated via HHG-based methods at lower photon energy and with larger driving intensity\cite{Nature.482.68}. Furthermore, it is realistic to assume that both the ion density and the sample length can be increased, using either EBITs\cite{BernittNature} or Paul traps\cite{PhysRevLett.86.1994}, in the case of lowly charged ions such as \mbox{Be$^{2+}$}. This would yield a further improvement in the output power per comb line.

Alternatively, experimental settings for the generation of the cloud of ions based on gas discharge or photoionization by an x-ray pre-pulse\cite{PhysRevLett.107.233001, Rohringer.Nature.481.2012} may lead to higher ion densities, with an improvement in the signal-to-background ratio. However, this would also result in a less stable ion cloud and one would have to ensure that a constant density of ions and atoms is encountered by the train of pulses from the optical-frequency-comb laser.
}

% The peaks of the predicted \mbox{x-ray} comb lie on top of the background power due to the transmitted \mbox{x-ray} FEL pulse. Within the spectral width $1/\tilde{\tau}_a$ of a single \mbox{x-ray}-comb peak, this corresponds to a power of
% $$I_{\mathrm{X1}} \mathscr{A}_{\mathrm{X1}}\, \frac{\tau_{\mathrm{X1}} }{\tilde{\tau}_a}\,\frac{1/\tilde{\tau}_a}{\Delta\omega_{\mathrm{X1}}} = 25\,\mathrm{nW}.$$
% The presence of the predicted \mbox{x-ray} comb could be detected by subtracting this background term.

\bibliographystyle{naturemag}